\documentclass[10pt, final, twocolumn]{IEEEtran}
\usepackage{amsmath,amssymb,amsfonts}
\usepackage{amsthm}
\usepackage{array}
\usepackage{algorithm}
\usepackage{algorithmic}
\usepackage{acronym}
\usepackage{bbm}
\usepackage{bm}
\usepackage{caption} 
\usepackage{cite}
\usepackage{colortbl}
\usepackage{comment}
\usepackage{enumerate}
\usepackage{float}
\usepackage{graphicx}
\usepackage{hhline}
\usepackage{hyperref}
\usepackage{multirow}
\usepackage{orcidlink}
\usepackage{pgfplots}
\pgfplotsset{compat=1.18}
\usepackage{psfrag}
\usepackage{stfloats}
\usepackage{subfig}
\usepackage{tabularx}
\usepackage{textcomp}
\usepackage{tikz}
\usepackage{url}
\usepackage{verbatim}
\usepackage{xcolor}
\usepackage{xparse}
\usepackage{soul}

\begin{document}

\title{Scalable RIS-Aided Beamforming Strategies for Near-Field MU-MISO via Multi-Antenna Feeder}

\author{\IEEEauthorblockN{G. Torcolacci,~\IEEEmembership{Graduate Student Member,~IEEE}, M. Schellmann, and D. Dardari,~\IEEEmembership{Fellow,~IEEE}} }

\markboth{Scalable RIS-Aided Beamforming Strategies for Near-Field MU-MISO via Multi-Antenna Feeder}
{Torcolacci \MakeLowercase{\textit{et al.}}}

\maketitle

\begin{abstract}
        This paper investigates a modular beamforming framework for reconfigurable intelligent surface (RIS)-aided multi-user (MU) communications in the near-field regime, built upon a novel antenna architecture integrating an active multi-antenna feeder (AMAF) array with a transmissive RIS (T-RIS), referred to as AT-RIS. This decoupling enables coordinated yet independently configurable designs in the AMAF and T-RIS domains, supporting flexible strategies with diverse complexity-performance trade-offs. Several implementations are analyzed, including diagonal and non-diagonal T-RIS architectures, paired with precoding schemes based on focusing, minimum mean-square error, and eigenmode decomposition. Simulation results demonstrate that while non-diagonal schemes maximize sum-rate in scenarios with a limited number of \acp{UE} and high angular separability, they exhibit fairness and scalability limitations as \ac{UE} density increases. Conversely, diagonal T-RIS configurations, particularly the proposed focusing-based scheme with uniform feeder-side power allocation, offer robust, fair, and scalable performance with minimal channel state information. The findings emphasize the critical impact of \acp{UE}' angular separability and reveal inherent trade-offs among spectral efficiency, complexity, and fairness, positioning diagonal AT-RIS architectures as practical solutions for scalable near-field MU multiple-input single-output systems.
\end{abstract}


%
%



\acrodef{$P_{EM}$}{probability of emulation, or false alarm}
\acrodef{$P_{FA}$}{probability of false alarm}
\acrodef{$P_{MD}$}{probability of missed detection}
\acrodef{$P_{D}$}{probability of detection}

\acrodef{2D}{bi-dimesional}
\acrodef{3D}{three-dimesional}
\acrodef{5G}{5th generation}
\acrodef{6G}{6th Generation}
\acrodef{ACF}{autocorrelation function}
\acrodef{ACG}{automatic	gain control}
\acrodef{ACI}{adjacent channel interference}
\acrodef{ACK}{acknowledge}
\acrodef{AcR}{autocorrelation receiver}
\acrodef{ADC}{analog-to-digital converter}
\acrodef{AF}{amplify \& forward}
\acrodef{AFL}{anchor-free localization}
\acrodef{AGNSS}{assisted-GNSS}
\acrodef{AGPS}{assisted GPS}
\acrodef{AI}{artificial intelligence}
\acrodef{AIC}{Akaike information criterion}
\acrodef{AMAF}{active multi-antenna feeder}
\acrodef{AO}{alternating optimization}
\acrodef{AOA}{angle-of-arrival}
\acrodef{AOD}{angle-of-departure}
\acrodef{AOT}{approximate optimum threshold}
\acrodef{AP}{access point}
\acrodef{API}{application programming interface}
\acrodef{ASK}{amplitude shift keying}
\acrodef{ASNR}{accumulated signal-to-noise ratio}
\acrodef{AT-RIS}{active transmissive reconfigurable intelligent surface}
\acrodef{AUB}{asymptotic union bound}
\acrodef{AWGN}{additive white Gaussian noise}
\acrodef{BAN}{body area network}
\acrodef{BAV}{balanced antipodal Vivaldi}
\acrodef{BCH}{Bose Chaudhuri Hocquenghem}
\acrodef{BEP}{bit error probability}
\acrodef{BER}{bit error rate}
\acrodef{BF}{brute force}
\acrodef{BFC}{block fading channel}
\acrodef{BIC}{Bayesian information criterion}
\acrodef{BLUE}{best linear unbiased estimator}
\acrodef{BPAM}{binary pulse amplitude modulation}
\acrodef{BPF}{bandpass filter}
\acrodef{BPPM}{binary pulse position modulation}
\acrodef{bps}{bits per second}
\acrodef{BPSK}{binary phase shift keying}
\acrodef{BPZF}{band-pass zonal filter}
\acrodef{BS}{base station}
\acrodef{BSC}{binary symmetric channel}
\acrodef{BTB}{Bellini-Tartara bound}
\acrodef{c.c.d.f.}{complementary cumulative distribution function}
\acrodef{c.d.f.}{cumulative distribution function}
\acrodef{CAD}{computer-aided design}
\acrodef{CAIC}{consistent Akaike information criterion}
\acrodef{CAP}{continuous aperture phased}
\acrodef{CCF}{cross correlation function}
\acrodef{CCI}{co-channel interference}
\acrodef{CD}{cooperative diversity}
\acrodef{CDMA}{code division multiple access}
\acrodef{CEOT}{channel ensemble optimum threshold}
\acrodef{CEP}{codeword error probability}
\acrodef{CFAR}{constant	 false alarm rate}
\acrodef{ch.f.}{characteristic function}
\acrodef{CH}{cluster head}
\acrodef{CIR}{channel impulse response}
\acrodef{CL}{centroid localization}
\acrodef{CM}{channel model}
\acrodef{CNR}{clutter-to-noise ratio}
\acrodef{CP}{ciclic prefix}
\acrodef{CPR}{channel pulse response}
\acrodef{CR}{channel response}
\acrodef{CRB}{Cram\'{e}r-Rao bound}
\acrodef{CRC}{cyclic redundancy check}
\acrodef{CRLB}{Cram\'{e}r-Rao lower bound}
\acrodef{CS}{clock skew}
\acrodef{CSCG}{circularly symmetric complex Gaussian}
\acrodef{CSI}{channel state information}
\acrodef{CSMA}{carrier sense multiple access}
\acrodef{CSS}{chirp spread spectrum}
\acrodef{CTS}{clear-to-send}
\acrodef{CW}{continuous wave}
\acrodef{DAA}{detect and avoid}
\acrodef{DAB}{digital audio broadcasting}
\acrodef{DAC}{digital-to-analog converter}
\acrodef{DBB}{digital base band}
\acrodef{DBPSK}{differential binary phase shift keying}
\acrodef{DCM}{dual-carrier modulation}
\acrodef{DDP}{detected direct path}
\acrodef{DF}{detect \& forward}
\acrodef{DFMS}{monopole dual feed stripline antenna}
\acrodef{DFT}{discrete Fourier transform}
\acrodef{DGPS}{differential GPS}
\acrodef{DLL}{delay-locked loop}
\acrodef{DNN}{deep neural network}
\acrodef{DOA}{direction of arrival}
\acrodef{DoD}{Department of Defense}
\acrodef{DoF}{degrees of freedom}
\acrodef{DP}{direct path}
\acrodef{DR}{detection rate}
\acrodef{DRT}{distance ratio test}
\acrodef{DS-SS}{direct-sequence spread-spectrum}
\acrodef{DS}{delay spread}
\acrodef{DTR}{differential transmitted-reference}
\acrodef{DTT}{Diffraction Tomography Theory}
\acrodef{DVB-H}{digital video broadcasting\,--\,handheld}
\acrodef{DVB-T}{digital video broadcasting\,--\,terrestrial}
\acrodef{e.m.}{electromagnetic}
\acrodef{ECC}{European Community Commission}
\acrodef{ED}{energy detector}
\acrodef{EDR}{energy detector receiver}
\acrodef{EFIM}{equivalent Fisher information matrix}
\acrodef{EIRP}{effective radiated isotropic power}
\acrodef{EKF}{extended Kalman filter}
\acrodef{KKT}{Karush–Kuhn–Tucker}
\acrodef{ELP}{equivalent low-pass}
\acrodef{EM}{electromagnetic}
\acrodef{EMCB}{extended Miller Chang bound}
\acrodef{EME}{minimum eigenvalue ratio detector}
\acrodef{EMI}{electromagnetic interference}
\acrodef{ENP}{estimated noise power}
\acrodef{ESA}{European Space Agency}
\acrodef{EU}{European Union}
\acrodef{EVD}{eigenvalue decomposition}
\acrodef{FAR}{false alarm rate}
\acrodef{FCC}{Federal Communications Commission}
\acrodef{FDMA}{frequency division multiple access}
\acrodef{FDMA}{frequency division multiple access}
\acrodef{DMA}{dynamic metasurface antenna}
\acrodef{FEC}{forward error correction}
\acrodef{FEC}{forward error correction}
\acrodef{FFD}{full function device}
\acrodef{FFR}{full function reader}
\acrodef{FF}{far-field}
\acrodef{FFT}{fast Fourier transform}
\acrodef{FG}{factor graph}
\acrodef{FH-SS}{frequency-hopping spread-spectrum}
\acrodef{FH}{frequency-hopping}
\acrodef{FIM}{Fisher information matrix}
\acrodef{FLL}{Frequency-locked loop}
\acrodef{FPGA}{field programmable gate array}
\acrodef{FS}{frame synchronization}
\acrodef{FT}{Fourier Transform}
\acrodef{GA}{Gaussian approximation}
\acrodef{GD}{gradient descent}
\acrodef{GDOP}{geometric dilution of precision}
\acrodef{GLR}{generalized likelihood ratio}
\acrodef{GLRT}{generalized likelihood ratio test}
\acrodef{GML}{generalized maximum likelihood}
\acrodef{GPRS}{general packet radio service}
\acrodef{GPS}{global positioning system}
\acrodef{HAP}{high altitude platform}
\acrodef{HCRB}{hybrid Cram\'{e}r-Rao bound}
\acrodef{HDSA}{high-definition situation-aware}
\acrodef{Hi-RADIAL}{High-accuracy RAdio Detection, Identification, And Localization}
\acrodef{HMIMO}{holographic multiple-input multiple-output}
\acrodef{HMM}{hidden Markov model}
\acrodef{HPA}{high-power amplifier}
\acrodef{HPBW}{half power beam width}
\acrodef{HW}{hardware}
\acrodef{i.i.d.}{independent, identically distributed}
\acrodef{ICT}{information and communication technologies}
\acrodef{IE}{informative element}
\acrodef{IEEE}{Institute of Electrical and Electronics Engineers}
\acrodef{IF}{intermediate frequency}
\acrodef{IFFT}{inverse fast Fourier transform}
\acrodef{IMF}{ideal matched filter}
\acrodef{IMU}{inertial measurement unit}
\acrodef{INR}{interference-to-noise ratio}
\acrodef{INS}{inertial navigation system}
\acrodef{IoT}{Internet of things}
\acrodef{IIoT}{industrial Internet of things}
\acrodef{INS}{inertial navigation system}
\acrodef{IR-UWB}{impulse radio UWB}
\acrodef{IR}{impulse radio}
\acrodef{IRI}{inter-reader interference}
\acrodef{IRS}{intelligent reflecting surface} 
\acrodef{ISAC}{integrated sensing and communications}
\acrodef{ISI}{inter-symbol interference} 
\acrodef{isi}{intra-symbol interference} 
\acrodef{ISM}{industrial, scientific and medical}
\acrodef{ISNR}{interference-plus-signal-to-noise-ratio}
\acrodef{ISP}{inverse scattering problem}
\acrodef{IT}{interference temperature}
\acrodef{ITC}{information theoretic criteria}
\acrodef{JBSF}{jump back and search forward}
\acrodef{JF}{just forward}
\acrodef{KF}{Kalman filter}
\acrodef{KKT}{Karush–Kuhn–Tucker}
\acrodef{LDC}{low duty cycle}
\acrodef{LDPC}{low density parity check}
\acrodef{LEO}{localization error outage}
\acrodef{LG}{Laguerre-Gaussian}
\acrodef{LIS}{large intelligent surface}
\acrodef{LLR}{log-likelihood ratio}
\acrodef{LLRT}{log-likelihood ratio test}
\acrodef{LRT}{likelihood ratio test}
\acrodef{LNA}{low-noise amplifier}
\acrodef{LOS}{line-of-sight}
\acrodef{LRT}{likelihood ratio test}
\acrodef{LS}{least square}
\acrodef{LS}{least squares}
\acrodef{M-PSK}{$M$-ary phase shift keying}
\acrodef{M-QAM}{$M$-ary quadrature amplitude modulation}
\acrodef{m.g.f.}{moment generating function}
\acrodef{MAC}{medium access control}
\acrodef{MAE}{mean absolute error}
\acrodef{MAI}{multiple access interference}
\acrodef{MAN}{metropolitan area network}
\acrodef{MAP}{maximum a posteriori}
\acrodef{MB-OFDM}{multi-band OFDM}
\acrodef{MB-UWB}{multi-band UWB}
\acrodef{MB}{multi-band}
\acrodef{MC}{multi-carrier}
\acrodef{MCB}{Miller Chang bound}
\acrodef{MCRB}{modified Cram\'{e}r-Rao bound}
\acrodef{MDD}{minimum distance distribution}
\acrodef{MDL}{minimum description length}
\acrodef{MF}{matched filter}
\acrodef{MGF}{moment generating function}
\acrodef{MI}{mutual information}
\acrodef{MIMO}{multiple-input multiple-output}
\acrodef{MISO}{multiple-input single-output}
\acrodef{ML}{maximum likelihood}
\acrodef{MM}{min-max}
\acrodef{MME}{maximum-minimum eigenvalue ratio detector}
\acrodef{mmWave}{millimeter wave}
\acrodef{MMSE}{minimum mean-square error}
\acrodef{MPC}{multipath component}
\acrodef{MRC}{maximal ratio combiner}
\acrodef{MRT}{maximum ratio transmission}
\acrodef{MS}{mobile station}
\acrodef{MSB}{most significant bit}
\acrodef{MSE}{mean squared error}
\acrodef{NMSE}{normalized mean squared error}
\acrodef{MSK}{minimum shift keying}
\acrodef{MU}{multi-user}
\acrodef{MUI}{multi-user interference}
\acrodef{MUR}{multistatic radar}
\acrodef{MVU}{minimum variance unbiased}
\acrodef{MZZB}{modified Ziv-Zakai bound}
\acrodef{NB}{narrowband}
\acrodef{NBI}{narrowband interference}
\acrodef{NEO}{navigation error outage}
\acrodef{NFER}{near-Þeld electromagnetic ranging}
\acrodef{NF}{near-field}
\acrodef{NFF}{near-field focused}
\acrodef{NL}{nonlinear}
\acrodef{NLOS}{non-line-of-sight}
\acrodef{NP}{Neyman-Pearson}
\acrodef{NTIA}{National Telecommunications and Information Administration}
\acrodef{NTP}{network time protocol}
\acrodef{OAM}{orbital angular momentum} 
\acrodef{OC}{optimum combining}
\acrodef{OFDM}{orthogonal frequency division multiplexing}
\acrodef{OOK}{on-off keying}
\acrodef{OP}{outage probability}
\acrodef{OT}{optimum threshold}
\acrodef{OTA}{over-the-air}
\acrodef{P-Max}{$P$-Max}  
\acrodef{p.d.f.}{probability density function}
\acrodef{p.m.f.}{probability mass function}
\acrodef{PA}{power amplifier}
\acrodef{PAM}{pulse amplitude modulation}
\acrodef{PAN}{personal area network}
\acrodef{PAR}{peak-to-average ratio}
\acrodef{P-CRLB}{Posterior Cramer-Rao Lower Bound}
\acrodef{PCA}{principal component analysis}
\acrodef{PD}{probability of detection}
\acrodef{PDP}{power delay profile}
\acrodef{PE}{probability of emulation}
\acrodef{PEB}{principal eigenmode beamforming}
\acrodef{PEC}{perfect electric conductor}
\acrodef{PEP}{packet error probability}
\acrodef{PF}{particle filter}
\acrodef{PFA}{probability of false alarm}
\acrodef{PHY}{physical layer}
\acrodef{PL}{path-loss}
\acrodef{PLL}{phase-locked loop}
\acrodef{PMD}{probability of missed detection}
\acrodef{PN}{pseudo-noise}
\acrodef{PSF}{point spread function}
\acrodef{ppm}{part-per-million}
\acrodef{PPM}{pulse position modulation}
\acrodef{PR}{pseudo-random}
\acrodef{PRake}{partial rake}
\acrodef{PRF}{pulse repetition frequency}
\acrodef{PRP}{pulse repetition period}
\acrodef{PSD}{power spectral density}
\acrodef{PSEP}{pairwise synchronization error probability}
\acrodef{PSNR}{peak signal to noise ratio}
\acrodef{PSK}{phase shift keying}
\acrodef{PSVD}{product singular value decomposition}
\acrodef{PSWF}{prolate spheroidal wave function}
\acrodef{PU}{primary user}
\acrodef{QAM}{quadrature amplitude modulation}
\acrodef{QoS}{quality of service}
\acrodef{QPSK}{quadrature phase shift keying}
\acrodef{R.V.}{random variable}
\acrodef{RADAR}{radar}
\acrodef{RCS}{radar cross section}
\acrodef{RDL}{"random data limit"}
\acrodef{REM}{radio environment map}
\acrodef{REO}{ranging error outage}
\acrodef{RF}{radio-frequency}
\acrodef{RFID}{radio-frequency identification}
\acrodef{RFR}{reduced function reader}
\acrodef{RFT}{reduced function tag}
\acrodef{RII}{ranging information intensity}
\acrodef{RIS}{reconfigurable intelligent surface}
\acrodef{rms}{root mean square}
\acrodef{RMSE}{root-mean-square error}
\acrodef{ROC}{receiver operating characteristic}
\acrodef{ROI}{region of interest}
\acrodef{RRC}{root raised cosine}
\acrodef{RSN}{radar sensor network}
\acrodef{RSS}{received signal strength}
\acrodef{RSSI}{received signal strength indicator}
\acrodef{RTLS}{real time locating systems}
\acrodef{RTT}{round-trip time}
\acrodef{S-V}{Saleh-Valenzuela}
\acrodef{SA}{simulated annealing}
\acrodef{SaG}{stop-and-go}
\acrodef{SAR}{synthetic aperture radar}
\acrodef{SBS}{serial backward search}
\acrodef{SBSMC}{serial backward search for multiple clusters}
\acrodef{SCM}{supply chain management}
\acrodef{SCR}{signal-to-clutter ratio}
\acrodef{SEP}{symbol error probability}
\acrodef{SIS}{small intelligent surface}
\acrodef{SFD}{start frame delimiter}
\acrodef{SIM}{stacked intelligent metasurface}
\acrodef{SIMO}{single-input multiple-output}
\acrodef{SINR}{signal-to-interference plus noise ratio}
\acrodef{SIR}{signal-to-interference ratio}
\acrodef{SISO}{single-input single-output}
\acrodef{SNR}{signal-to-noise ratio}
\acrodef{SoC}{system on chip}
\acrodef{SoO}{signal of opportunity}
\acrodef{SoP}{system on package}
\acrodef{SOT}{sub-optimum threshold}
\acrodef{SPAWN}{sum-product algorithm over a wireless network}
\acrodef{SPEB}{squared position error bound}
\acrodef{SPMF}{single-path matched filter}
\acrodef{SPP}{spiral phase plate}
\acrodef{SQNR}{signal-to-quantization-noise ratio}
\acrodef{SRE}{smart radio environment}
\acrodef{SS}{spread spectrum}
\acrodef{ST}{simple thresholding}
\acrodef{SU}{single-user}
\acrodef{SVD}{singular value decomposition}
\acrodef{SW}{software}
\acrodef{SW}{sync word}
\acrodef{TDE}{time delay estimation}
\acrodef{TDL}{tapped delay line}
\acrodef{TDMA}{time division multiple access}
\acrodef{TDOA}{time difference-of-arrival}
\acrodef{TH}{time-hopping}
\acrodef{THz}{terahertz}
\acrodef{TNR}{threshold-to-noise ratio}
\acrodef{TOA}{Time-of-arrival}
\acrodef{TOF}{time-of-flight}
\acrodef{TPC}{transmit power control}
\acrodef{TR}{transmitted-reference}
\acrodef{T-RIS}{transmissive reconfigurable intelligent surface}
\acrodef{TS}{tabu search}
\acrodef{TSVD}{truncated singular value decomposition}
\acrodef{TV}{total variation denoising}
\acrodef{UAV}{unmanned aerial vehicle}
\acrodef{UB}{union bound}
\acrodef{UCA}{uniform circular array}
\acrodef{UDP}{undetected direct path}
\acrodef{UE}{User Equipment}
\acrodef{UHF}{ultra-high frequency}
\acrodef{ULA}{uniform linear array}
\acrodef{UPA}{uniform planar array}
\acrodef{ULP}{user location protocol}
\acrodef{UMP}{uniformly most powerful}
\acrodef{UMPI}{uniformly most powerful invariant}
\acrodef{URA}{uniform rectangular array}
\acrodef{UT}{user terminal}
\acrodef{UTC}{coordinated universal time}
\acrodef{UTM}{universal transverse Mercator}
\acrodef{UTRA}{UMTS terrestrial radio access}
\acrodef{UAV}{unmanned aerial vehicle}
\acrodef{UUV}{unmanned underwater vehicle}
\acrodef{UWB}{ultrawide-band}
\acrodef{UWBcap}[UWB]{Ultrawide band}
\acrodef{VFIL}{virtual force iterative localization}
\acrodef{VGA}{variable-gain amplifier}
\acrodef{VNA}{vector network analyzer}
\acrodef{WAF}{wall attenuation factor}
\acrodef{WB}{wideband}
\acrodef{WBI}{wideband interference}
\acrodef{WCL}{weighted centroid localization}
\acrodef{WED}{wall extra delay}
\acrodef{WiMAX} {worldwide interoperability for microwave access}
\acrodef{WLAN}{wireless local area network}
\acrodef{WLS}{weighted least squares}
\acrodef{WMAN}{wireless metropolitan area network}
\acrodef{WMMSE}{weighted minimum mean-square error}
\acrodef{WPAN}{wireless personal area networks}
\acrodef{WRAPI}{wireless research application programming interface}
\acrodef{WSN}{wireless sensor network}
\acrodef{WSR}{wireless sensor radar}
\acrodef{WSS}{wide-sense stationary}
\acrodef{WWB}{Weiss-Weinstein bound}
\acrodef{WWLB}{Weiss-Weinstein lower bound}
\acrodef{ZZB}{Ziv-Zakai bound}
\acrodef{ZZLB}{Ziv-Zakai lower bound}
\acrodef{XL-MIMO}{extremely large-scale multiple-input multiple-output}


%

\newcommand{\rect}[1] {\text{rect} \left ({#1} \right )}
\newcommand{\sinc}[1] {\text{sinc} \left ({#1} \right )}
\newcommand{\argmax}[1]{\underset{{#1}}{\operatorname{argmax}}}
\newcommand{\argmin}[1]{\underset{{#1}}{\operatorname{argmin}}}
\newcommand{\E}[1] {\mathbb{E}\left\{#1\right\}}
\newcommand{\Var}[1] {\mathbb{V}\left\{#1\right\}}
\newcommand{\Real}[1]{\Re^{#1}}
\newcommand{\floor}[1] {f \left ({#1} \right )}
\def\erfc{{\text{erfc}}}
\def\erf{{\text{erf}}}
\def\inverfc{{\text{inverfc}}}
\newcommand{\rank}{{\rm rank}}
\newcommand{\diag}{{\rm diag}}
\newcommand{\tr}{{\rm tr}}
\newcommand{\degree}{\ensuremath{^\circ}}
\newcommand{\ra}{\rightarrow}
\newcommand{\rf}{\leftarrow}
\newcommand{\cn}{{\mathcal{CN}}} 
\newcommand{\Variance}[1]{\operatorname{Var} \left({#1} \right)}

\newcommand{\SNR}{\text{SNR}}
\newcommand{\TNR}{\mathsf{TNR}}
\newcommand{\sigmaN} {\sigma_{\text{N}}}
\newcommand{\MSE} {\text{MSE}}

\newcommand{\bolda}{{\bf a}}
\newcommand{\boldb}{{\bf b}}
\newcommand{\boldbeta}{{\boldsymbol{\beta}}}
\newcommand{\boldc}{{\bf c}}
\newcommand{\boldd}{{\bf d}}
\newcommand{\bolde}{{\bf e}}
\newcommand{\boldf}{{\bf f}}
\newcommand{\boldg}{{\bf g}}
\newcommand{\boldh}{{\bf h}}
\newcommand{\boldi}{{\bf i}}
\newcommand{\boldj}{{\bf j}}
\newcommand{\boldk}{{\bf k}}
\newcommand{\boldl}{{\bf l}}
\newcommand{\boldm}{{\bf m}}
\newcommand{\boldn}{{\bf n}}
\newcommand{\boldo}{{\bf o}}
\newcommand{\boldp}{{\bf p}}
\newcommand{\boldq}{{\bf q}}
\newcommand{\boldr}{{\bf r}}
\newcommand{\bolds}{{\bf s}}
\newcommand{\boldsp} {{\bf s}^{\prime}}
\newcommand{\boldt}{{\bf t}}
\newcommand{\boldu}{{\bf u}}
\newcommand{\boldv}{{\bf v}}
\newcommand{\boldw}{{\bf w}}
\newcommand{\boldx}{{\bf x}}
\newcommand{\boldy}{{\bf y}}
\newcommand{\boldz}{{\bf z}}

\newcommand{\boldA}{{\bf A}}
\newcommand{\boldB}{{\bf B}}
\newcommand{\boldC}{{\bf C}}
\newcommand{\boldD}{{\bf D}}
\newcommand{\boldE}{{\bf E}}
\newcommand{\boldF}{{\bf F}}
\newcommand{\boldG}{{\bf G}}
\newcommand{\boldH}{{\bf H}}
\newcommand{\boldI}{{\bf I}}
\newcommand{\boldJ}{{\bf J}}
\newcommand{\boldK}{{\bf K}}
\newcommand{\boldL}{{\bf L}}
\newcommand{\boldM}{{\bf M}}
\newcommand{\boldN}{{\bf N}}
\newcommand{\boldO}{{\bf O}}
\newcommand{\boldP}{{\bf P}}
\newcommand{\boldQ}{{\bf Q}}
\newcommand{\boldR}{{\bf R}}
\newcommand{\boldS}{{\bf S}}
\newcommand{\boldT}{{\bf T}}
\newcommand{\boldU}{{\bf U}}
\newcommand{\boldV}{{\bf V}}
\newcommand{\boldW}{{\bf W}}
\newcommand{\boldX}{{\bf X}}
\newcommand{\boldY}{{\bf Y}}
\newcommand{\boldZ}{{\bf Z}}

\newcommand{\stx}{S_{\text{T}}}
\newcommand{\srx}{S_{\text{R}}}
\newcommand{\lt}{L_{\text{T}}}
\newcommand{\lr}{L_{\text{R}}}
\newcommand{\rhot}{\rho_{\text{T}}}
\newcommand{\phit}{\varphi_{\text{T}}}
\newcommand{\rhor}{\rho_{\text{R}}}
\newcommand{\phir}{\varphi_{\text{R}}}
\newcommand{\Pt}{P_{\text{T}}}
\renewcommand{\Pr}{P_{\text{R}}}   
\newcommand{\Nt}{N_{\text{T}}}
\newcommand{\Nr}{N_{\text{R}}}
\newcommand{\Ns}{N_{\text{S}}}
\newcommand{\Gt}{{\bf G_{\text{T}}}}
\newcommand{\Gr}{{\bf G_{\text{R}}}}
\newcommand{\lambdaA}{\lambda_{\boldA}}
\newcommand{\vA}{\mathbf{v}_{\mathbf{A}}}
\newcommand{\gammastar}{{\boldsymbol{\gamma}}^\star}
\newcommand{\Ndof}{N_{\text{DOF}}}
\newcommand{\bbeta}{{\bm {\beta}}}
\newcommand{\bchi}{{\bm{\chi}}}
\newcommand{\bsigma}{{\bm{\Sigma}}}
\newcommand{\blambda}{{\bm{\Lambda}}}
\newcommand{\invsigma}{{\bm{\Sigma}}^{-1}}
\newcommand{\pinvsigma}{{\bm{\Sigma}}^{\dagger}}
\newcommand{\tbw}{\tilde{\boldw}}
\newcommand{\tbsigma}{\tilde{\bm{\Sigma}}}
\newcommand{\hbbeta}{{\hat{\boldsymbol{\beta}}}}
\newcommand{\betan}{\beta_n}
\newcommand{\xin}{\xi_n}
\newcommand{\sumn}{\sum_{n=1}^\infty}
\newcommand{\tildexn}{\tilde{x}_n}
\newcommand{\xn}{x_n}
\newcommand{\xns}{x_n^\star}
\newcommand{\lambdas}{\lambda^\star}
\newcommand{\fz}{f_0}
\newcommand{\fo}{f_1}

\newcommand{\Ht}{\boldH_{\text{T}}}
\newcommand{\Hr}{\boldH_{\text{R}}}

\newcommand{\fig}[1]{Fig.~\ref{#1}}
\newcommand{\sect}[1]{Sec.~\ref{#1}}
\newcommand{\apd}[1]{Appendix~\ref{#1}}
\newcommand{\eq}[1]{(\ref{#1})}

\newcommand{\gtij}{g_{\text{T}, i, j}}
\newcommand{\Thetanm}{\boldsymbol{\Theta}_{n,m}}

\newcommand{\rt}{R_{\text{T}}}

\maketitle
\acresetall

\section{Introduction}

\IEEEPARstart{T}{he} ever-increasing demand for higher capacity, lower latency, and enhanced spatial resolution in future wireless networks is steering the design of \ac{6G} systems toward higher frequency bands and extremely large antenna arrays operating in the radiative near-field region. Among the technologies supporting this evolution, \ac{XL-MIMO} systems have emerged as a key enabler, leveraging massive numbers of antennas and near-field propagation to maximize spectrum efficiency, hence unlock unprecedented spatial multiplexing capabilities. Nevertheless, the practical deployment of fully active \ac{XL-MIMO} arrays remains highly challenging due to the excessive hardware complexity, cost, and power consumption associated with assigning a dedicated \ac{RF} chain to each antenna element. To overcome these limitations, several alternative antenna architectures have been explored, aiming to strike a balance between performance and hardware efficiency. Notable examples include \acp{DMA} \cite{zhang2021beam}, hybrid analog-digital beamforming architectures \cite{sohrabi2017hybrid}, and holographic beamforming based on programmable metasurfaces \cite{li2023near,an2023tutorial}.

A novel antenna architecture has recently emerged, combining an \ac{AMAF} with a \ac{RIS} \cite{clemente2024hybrid, milbrandt20232}, which can be configured to operate in either reflective or transmissive mode depending on the design requirements. In this work, we specifically focus on the transmissive configuration, commonly referred to as \ac{T-RIS}. This new antenna paradigm, referred to in this work as \ac{AT-RIS}, provides an energy-efficient and scalable solution for \ac{MU} wireless communications by leveraging two key advantages. On one hand, the near-field \ac{EM} interaction between the \ac{AMAF} and the \ac{T-RIS} enables efficient space-feeding, i.e., wireless energy transfer via free-space propagation, allowing the excitation of multiple spatial modes across the \ac{T-RIS} surface, thereby supporting spatial multiplexing without requiring dedicated \ac{RF} chains. On the other hand, the programmable nature of the \ac{T-RIS} allows dynamic control over the impinging wavefront, enabling real-time manipulation of key \ac{EM} beam properties such as direction, focal point, and shape. 

Interestingly, the \ac{AT-RIS} architecture offers several compelling advantages over conventional active and hybrid array systems. First, it significantly reduces hardware complexity and power consumption by minimizing the number of active \ac{RF} chains, while partially offloading signal processing to the \ac{EM} domain. Second, the metasurface-based \ac{T-RIS} offers high reconfigurability, supporting dynamic beamforming, focusing, and angular steering via tunable meta-atoms. Third, unlike traditional hybrid arrays that rely on analog feeding networks, often subject to insertion losses and limited scalability~\cite{abdelrahman2017analysis}, the \ac{AT-RIS} adopts a space-fed design, where the \ac{T-RIS} is wirelessly illuminated by a spatially separated active source. This decoupled feeding approach eliminates physical interconnections, thereby potentially enhancing radiation efficiency and relaxing deployment constraints. Furthermore, when the excitation pattern of the \ac{AMAF} and the configuration of the \ac{T-RIS} are jointly optimized, the architecture allows for partial mitigation of classical aperture-related impairments, such as spillover and scanning losses, hence enabling enhanced control over beam properties including beam squinting and grating lobes~\cite{lau2012reconfigurable,wang2022beam}. However, the extent to which these impairments can be compensated depends on the specific metasurface design, available \ac{DoF}, and the operating scenario. Finally, the use of large passive apertures in the \ac{T-RIS} enables high effective antenna gain and wide angular coverage, all while preserving low hardware complexity and cost, i.e., benefits that are challenging to achieve with conventional phased array or hybrid architectures \cite{demmer2023hybrid}.
    
\subsection{State of the Art} 

Recent studies have investigated the fundamental properties and beamforming potential of the emerging \ac{AT-RIS} architecture. In particular, works such as \cite{TiwCai:C23, TiwCai:C22, TiwCai:C24} focus on the case where both the \ac{AMAF} and \ac{T-RIS} are implemented as \acp{ULA}, characterizing the near-field propagation matrix between the two entities, which is a critical step for enabling joint beamforming optimization. Subsequently, the same authors extended their investigations by considering the case where both \ac{T-RIS} and \ac{AMAF} are \acp{UPA} \cite{TiwCai:J23, TiwCai2:C23}. They develop a single-beam precoding strategy based on \ac{SVD}, referred to as the \ac{PEB} design. This approach aims to maximize power transfer and beamforming efficiency between the \ac{AMAF} and the \ac{T-RIS}, thus effectively shaping the radiated field over the ground plane in narrowband systems. The \ac{PEB} scheme demonstrates robust performance over wide signal bandwidths and offers near-optimal beamforming in \ac{MU} scenarios with limited inter-\ac{UE} interference. However, its extension to the multiple \acp{UE} case relies on vertically stacking independent \ac{AT-RIS} modules, one per \ac{UE}, since the system inherently generates a single spatial beam at the feeder side. Although effective in specific propagation scenarios, this solution suffers from poor scalability and becomes impractical in dense or dynamic deployments. In this context, designing a single, adaptive \ac{AT-RIS} module capable of serving multiple \acp{UE} via programmable radiation patterns emerges as a promising and resource-efficient alternative. Exploring this design space and quantifying the associated trade-offs in terms of performance, fairness, and complexity constitutes the central focus of this work.

In addition, the work in~\cite{jamali2020intelligent} offers a comprehensive overview and comparative analysis of emerging antenna architectures for \ac{MIMO} systems, contrasting such \ac{T-RIS}-assisted antenna designs with conventional fully digital, hybrid, and lens-based arrays. Focusing specifically on the \ac{AT-RIS} configuration, the authors examine the impact of different illumination strategies, including full, partial, and separate illumination, where the latter refers to a special case of partial illumination in which each section of the \ac{T-RIS} is excited by a dedicated active antenna. Within this framework, two precoding schemes are proposed: one that maximizes \ac{MI}, and another based on orthogonal matching pursuit (OMP), i.e., a greedy algorithm that iteratively selects basis vectors from a given dictionary to approximate the optimal precoding vector. The results reveal that the OMP-based approach performs better in sparse scattering environments, whereas the MI-based method offers superior spectral efficiency in rich multipath scenarios. These findings underscore the relevance of adapting precoding strategies to channel conditions, while also highlighting trade-offs between computational complexity and achievable performance.

Furthermore, the study in~\cite{buzzi2022approaching} analyzes \ac{AT-RIS}-based systems comprising a compact \ac{AMAF} array in close proximity to the \ac{T-RIS}, considering both passive and active \ac{T-RIS} implementations. In the active variant, the \ac{T-RIS} elements not only apply phase shifts but also amplify the incoming signals. A central insight is that favorable propagation and channel hardening conditions depend more critically on the number of active elements at the \ac{AMAF} than on the physical size of the \ac{T-RIS}. The analysis demonstrates that the additional spatial \ac{DoF} introduced by the metasurface can be exploited to enhance inter-\ac{UE} channel orthogonality, thereby improving \ac{MU} performance. These findings highlight the critical role of properly balancing active and passive components in \ac{AT-RIS} architectures to fully exploit the spatial diversity enabled by large passive apertures.

\subsection{Main Contributions}
Overall, the aforementioned studies lay the groundwork for exploring this emerging antenna paradigm, offering valuable insights into its capabilities and limitations. Nonetheless, the full potential of \ac{AT-RIS} architectures remains largely unexplored in the context of \ac{MU} deployments. While prior works predominantly focus on specific \ac{T-RIS} implementations, e.g., assuming diagonal phase-only configurations, and address beamforming design for \ac{SU} scenarios, their generalization to dense \ac{MU} settings poses significant challenges. In particular, unlike prior solutions such as \cite{TiwCai2:C23}, which rely on per-\ac{UE} modular deployments under far-field assumptions, the joint optimization of radiation patterns and power allocation strategies that efficiently serve multiple \acp{UE} with a single \ac{AT-RIS} module across both near- and far-field regimes remains an open research problem. 

In this work, we build upon these foundational contributions by conducting a comparative analysis across multiple \ac{T-RIS} architectures, i.e., diagonal and non-diagonal, and \ac{AMAF}-side beamforming strategies, including dynamic power allocation among multiple excitation vectors. Our goal is to quantify the impact of these design choices on system-level performance in both near- and far-field regimes, ultimately providing practical guidelines for scalable and efficient \ac{MU} deployments. In this regard, the main contributions of the paper can be summarized as follows:

\begin{itemize}
    \item We develop a unified analytical and simulation framework for \ac{AT-RIS}-aided \ac{MU} \ac{MISO} systems operating in \ac{LOS} conditions, explicitly capturing the interplay between the active \ac{AMAF} and passive \ac{T-RIS} components. The model leverages near-field propagation to exploit spatial multiplexing capabilities at scale.

    \item We propose a suite of scalable beamforming strategies spanning from low-complexity focusing-based schemes to advanced \ac{MMSE}- and eigenmode-based designs. Each scheme jointly optimizes the \ac{AMAF} excitation and \ac{T-RIS} phase configuration, offering different trade-offs between performance and implementation complexity.

    \item We perform an extensive simulation-based evaluation across four critical dimensions: \ac{UE} angular spacing, link distance, power allocation method, and \ac{UE} density. The results characterize the operating regimes and robustness of each scheme under diverse and realistic deployment conditions.

    \item We demonstrate that diagonal \ac{T-RIS} architectures, when combined with appropriate beamforming and power allocation techniques, can achieve competitive performance compared to their non-diagonal counterparts, while ensuring significantly lower complexity. Importantly, they provide improved fairness in dense \ac{UE} scenarios, compared to more complex non-diagonal architectures.

    \item We extract actionable design insights for practical \ac{AT-RIS} deployments, illustrating how spatial \ac{DoF}, \ac{UE} geometry, and propagation regime shape system performance. These insights pave the way for future work on fairness-aware, energy-efficient, and robust optimization methodologies.
\end{itemize}

The remainder of the paper is organized as follows. Sec.~\ref{sec:SystemModel} introduces the system model, covering both single- and \ac{MU} communication scenarios. In Sec.~\ref{sec:ConfigStrategies}, we present a set of beamforming strategies for jointly configuring the \ac{AMAF} and \ac{T-RIS} for sum rate maximization, considering both diagonal and non-diagonal architectures. Sec.~\ref{sec:NumericalResults} reports the corresponding numerical results and performance analysis. Finally, Sec.~\ref{sec:Conclusions} draws the main conclusions and outlines directions for future work.

\begin{figure*}[th!]
    \centering
    \includegraphics[width=0.65\textwidth, keepaspectratio, trim=0 0 0 0, clip]{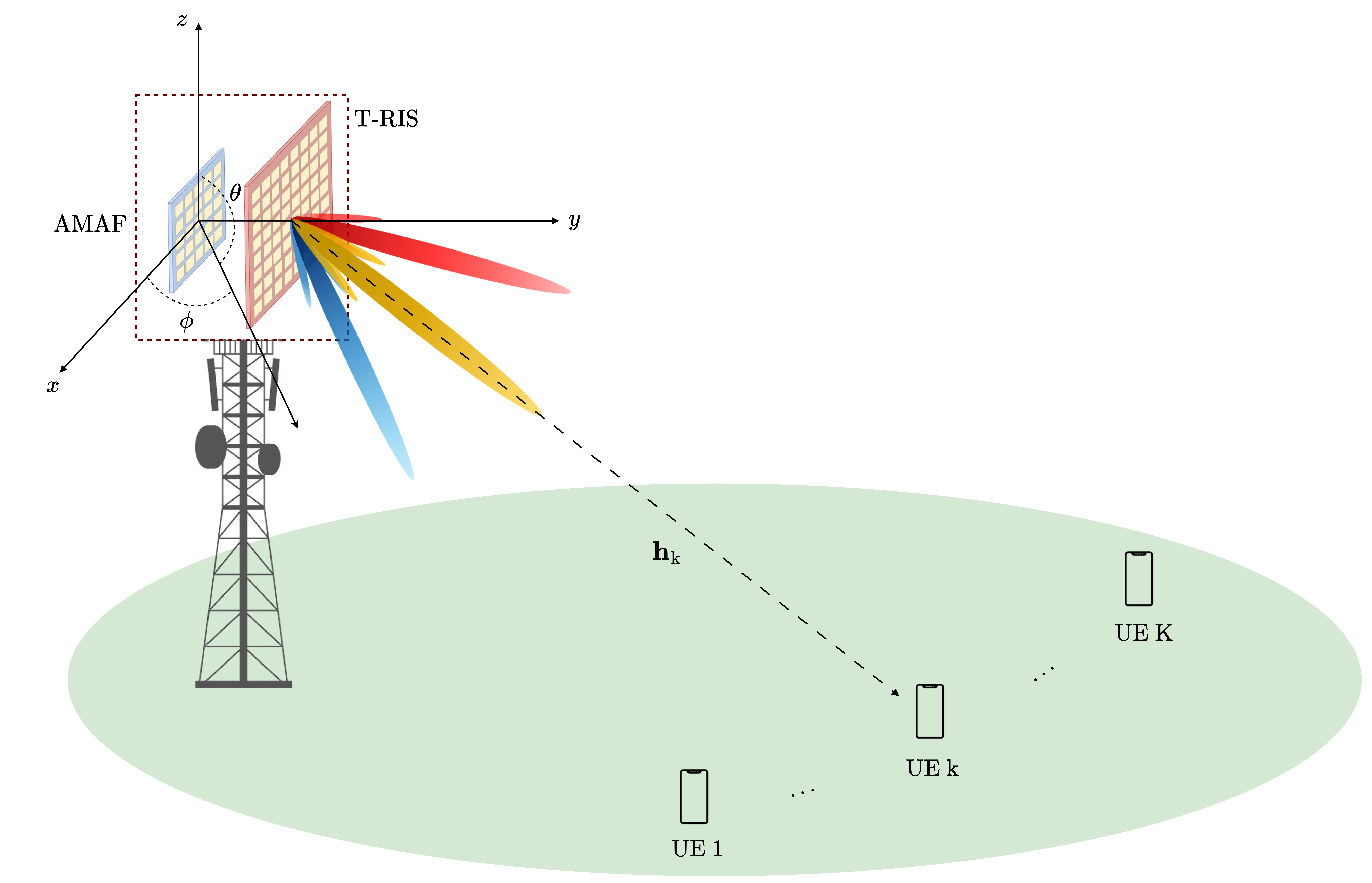}
    \caption{\ac{AT-RIS} antenna architecture for \ac{MU}-\ac{MISO} communications in \ac{LOS} free-space conditions.}
    \label{fig:scenario}
\end{figure*}

\subsection{Notation and Definitions}
Throughout this paper, we adopt the following notation. Vectors in \ac{3D} space are represented by lowercase bold letters, such as $\mathbf{x}$. Bold capital letters denote matrices, e.g., $\mathbf{X}$. The identity and zero matrices of size $N \times M$ are expressed as $\mathbf{I}_{N \times M}$ and $\mathbf{0}_{N \times M}$, respectively. The transpose of a matrix is indicated by $(\cdot)^T$, the Hermitian transpose by $(\cdot)^H$, the Moore-Penrose pseudoinverse by $(\cdot)^{\dagger}$, while its trace and its determinant are respectively denoted as $\tr\left(\cdot\right)$ and $\det\left(\cdot\right)$. The L2-norm of a vector $\mathbf{r}$ is denoted by $\|\mathbf{r}\|$, while the Frobenius norm of a matrix $\mathbf{X}$ is given by $\|\mathbf{X}\|_F$. Operator $\angle \boldx$ provides a vector containing the phases of each $\boldx$ element, while $\odot$ denotes the Hadamard, i.e., element-wise, product. The imaginary unit is represented by $\jmath$, while $\E{\cdot}$ is the expectation operator. Calligraphic letters are used to signify sets, such as $\mathcal{X}$, and a complex random vector $\mathbf{x} \sim \mathcal{CN}(\boldsymbol{\mu}, \boldsymbol{\Sigma})$ is distributed according to a complex normal distribution with mean $\boldsymbol{\mu}$ and covariance matrix $\boldsymbol{\Sigma}$. The notation $\text{diag}(\mathbf{x})$ refers to an operator that creates a diagonal matrix with the vector $\mathbf{x}$ on its main diagonal. Finally, the singular values of a matrix $\mathbf{A} \in \mathbb{C}^{N \times M}$ are denoted as $\xi_1(\mathbf{A}) \geq \xi_2(\mathbf{A}) \geq \ldots \geq \xi_K(\mathbf{A})$, where $K = \min(N, M)$. 

\section{Scenario and Problem Formulation}\label{sec:SystemModel}

Let us consider a transmitting \ac{BS} equipped with a \ac{AT-RIS} antenna that communicates with $K$ \acp{UE}, as depicted in Fig.~\ref{fig:scenario}. Specifically, the \ac{AMAF} array is a square \ac{UPA} composed of $\Nt$ active antenna elements located at $\boldp_{\text{T}, i} = \left[ x_{\text{T}, i},\, y_{\text{T}, i}, \, z_{\text{T}, i} \right]^T$, with $i \in \mathcal{N}_{\text{T}}=\{1,2, \ldots, \Nt\}$, and spaced of $\lambda/2$, which illuminates the \ac{T-RIS} with a signal with wavelength $\lambda$. Moreover, the \ac{T-RIS} has a square shape and is composed of $N$ unit cells (or meta-atoms), each of them located in $\boldp_{\text{RIS}, j} = \left[ x_{\text{RIS}, j},\, y_{\text{RIS}, j},\,z_{\text{RIS}, j} \right]^T$, with $j \in \mathcal{N}=\{1,2, \ldots, N\}$, and spaced of $\lambda/2$, while each of the $K$ \acp{UE} is equipped with a single antenna element placed at $\boldp_{\text{R}}^{(k)} = \left[ x_{\text{R}}^{(k)},\, y_{\text{R}}^{(k)},\,z_{\text{R}}^{(k)} \right]^T$, with $k \in \mathcal{K}=\{1,2,\ldots, K\}$.

We assume that the \ac{T-RIS} is located in the radiative near-field region of the \ac{AMAF} array, i.e., it holds \cite{balanis} 
\begin{equation}
    \label{eq:NFcond}
2D\sqrt{2N} \leq d \leq \frac{2\left(D \sqrt{2N}\right)^2}{\lambda} = d_{\text{FF}}\;,
\end{equation}
where $d$ is the distance between the \ac{AMAF} and \ac{T-RIS} centers and being $D$ the length of the \ac{T-RIS} diagonal. This definition corresponds to the array equivalent of the Fresnel region of an antenna \cite{bjornson2021primer}, allowing for the identification of the spatial area in which both amplitude and phase variations of the \ac{EM} spherical wavefronts must be considered when comparing local phases among antenna elements, despite the wave appearing locally planar at each element. 

In contrast, the propagation regime between the \ac{T-RIS} and each \ac{UE} depends on the deployment geometry and may vary across \acp{UE}. Specifically, the distance $d^{\prime}$ between the \ac{T-RIS} and the $k$th \ac{UE} may fall in either the near-field or far-field region of the \ac{T-RIS}, i.e., $d^{\prime} < d_{\text{FF}}$ or $d^{\prime} > d_{\text{FF}}$~\cite{balanis}. Nevertheless, to maintain generality and support arbitrary \ac{UE} locations, we consistently adopt a near-field channel model for all links, capturing both amplitude and phase variations of the spherical wavefronts.

\subsection{Single UE Communication}

We begin by analyzing the \ac{SU} transmission scenario, which serves as a building block for the more general \ac{MU} case discussed later. For notational consistency and ease of extension, we adopt the subscript $k\in \mathcal{K} $ to denote quantities associated with the served \ac{UE} $k$, though only a single \ac{UE} is considered in this section. Let $x_k \in \mathbb{C}$ be the transmitted symbol with $\mathbb{E}\left\{|x_k|^2\right\} = 1$. The received signal $y_k \in \mathbb{C}$ at the $k$th \ac{UE} is then given by
\begin{align}\label{eq:singleUEmode}
y_k = \boldh_k \boldsymbol{\Phi} \boldG \boldb_k x_k + w_k = \bar{\boldh}_k \boldb_k x_k + w_k  \, ,
\end{align}
where $\boldb_k =  \left[ b_1, \dots, b_i, \ldots, b_{\Nt}\right]^T \in\mathbb{C}^{\Nt \times 1}$ is the beamforming vector employed at the \ac{AMAF} for the $k$th \ac{UE} transmission, $\boldG = \{ g_{j,i}\} \in \mathbb{C}^{N \times \Nt}$ is the channel matrix describing the \ac{EM} waves propagation occurring between the \ac{AMAF} array and the \ac{T-RIS}, while $\boldsymbol{\Phi} \in \mathbb{C}^{N \times N}$ is the matrix characterizing the \ac{T-RIS} configuration and \ac{EM} processing applied to the impinging waveforms. We consider a transmissive metasurface that captures the incident signal from the side facing the \ac{AMAF} and re-emits it toward the \acp{UE} from the opposite side, applying element-wise amplitude and phase transformations specified by the matrix $\boldsymbol{\Phi}$. Moreover, we assume the \ac{T-RIS} to be lossless and passive, i.e., it redistributes the incident power without amplification, which corresponds to impose $ \boldsymbol{\Phi}^H \boldsymbol{\Phi} = \boldI_N$. Lastly, $\boldh_k = \left[ h_{1}^{(k)}, \ldots,  h_{j}^{(k)}, \ldots,  h_{N}^{(k)} \right] \in \mathbb{C}^{1 \times N}$ is the channel vector describing signal propagation between the \ac{T-RIS} and the $k$th \ac{UE}, and $w_k \sim \mathcal{CN}\left(0, \sigmaN^2 \right) \in \mathbb{C}$ is the \ac{AWGN}. For notational convenience, we introduce the $k$th \ac{UE} effective channel vector as $ \bar{\boldh}_k =   \boldh_k \boldsymbol{\Phi} \boldG \in \mathbb{C}^{1 \times \Nt}$.

By assuming \ac{LOS} free-space propagation, the elements of the channel matrix $\boldG$ can be computed as
\begin{equation}\label{eq:gcoeff}
    g_{j,i} = \frac{\lambda}{4 \pi d_{i,j}}\sqrt{G_{\text{T}}\left( \boldsymbol{\Theta}_{i,j}\right)} \,e^{- \jmath \frac{2 \pi}{\lambda}d_{i,j}} \;, 
\end{equation}
where $\left[d_{i,j},  \boldsymbol{\Theta}_{i,j}\right]=\left[d_{i,j}, \phi_{i,j}, \theta_{i,j}\right]$ represent the distance, azimuth angle, and elevation angle between the $(\boldp_{\text{T}, i},\boldp_{\text{RIS}, j} )$ pair, while $G_{\text{T}}\left( \boldsymbol{\Theta}_{i,j}\right)$ is the \ac{AMAF} transmitting beam pattern gain evaluated in the direction of departure $\boldsymbol{\Theta}_{i,j}$. Similarly, we can compute the channel coefficients for the RIS-$k$th \ac{UE} link $\boldh_k$ as per
\begin{align}\label{eq:hk_coeff}
  h_{j}^{(k)}  =\frac{\lambda}{4 \pi d_{j}^{(k)}} \sqrt{G_{\text{R}}\left( \boldsymbol{\Theta}_{j}^{(k)}\right)}\, e^{- \jmath \frac{2 \pi}{\lambda}d_{j}^{(k)}} \,, \; k \in \mathcal{K}\;,
\end{align}
with $\left[d_{j}^{(k)},  \boldsymbol{\Theta}_{j}^{(k)}\right]=\left[d_{j}^{(k)}, \phi_{j}^{(k)}, \theta_{j}^{(k)}\right]$ identifying the distance and angles between the $j$th \ac{T-RIS} unit cell and the $k$th \ac{UE}, and being $G_{\text{R}}\left( \boldsymbol{\Theta}_{j}^{(k)}\right)$ the receiving beam pattern gain evaluated in the direction of arrival $\boldsymbol{\Theta}_{j}^{(k)}$, seen from the \ac{UE} perspective. Notably, the proposed received signal model applies to both near-field and far-field propagation regimes by taking into account the precise distances and angles between each pair of antennas\footnote{This approach inherently captures the smooth and physically consistent transition between near- and far-field regions, without relying on simplified approximations.} \cite{torcolacci2024holographic}.

To facilitate the subsequent analysis, we define the \ac{SVD} of the channel matrix $\boldG$ between the \ac{AMAF} and the \ac{T-RIS} as
\begin{equation}\label{eq:G_svd}
\boldG \overset{\text{SVD}}{=} \boldU \boldsymbol{\Lambda} \boldV^H = \sum_{i=1}^{M} \xi_i \boldu_i \boldv_i^H,
\end{equation}
where $\boldsymbol{\Lambda} = \diag \left(\xi_1, \xi_2, \ldots, \xi_M\right) \in \mathbb{C}^{N \times \Nt}$ is a rectangular diagonal matrix containing the singular values of $\boldG$, with $M = \min (\Nt, N)$ and $\xi_i(\boldG) \in \mathbb{R}_0^+$, $\forall i \in \mathcal{M} = \{1, 2, \ldots, M\}$. The matrices $\boldU \in \mathbb{C}^{N \times N}$ and $\boldV \in \mathbb{C}^{\Nt \times \Nt}$ are unitary and contain the left and right singular vectors of $\boldG$, respectively. In particular, the columns $\boldu_i$ of $\boldU$ and $\boldv_i$ of $\boldV$ represent the eigenvectors of $\boldG \boldG^H$ and $\boldG^H \boldG$, respectively. Moreover, we define the matrix $\boldH = \left[\boldh_1, \ldots, \boldh_k, \ldots, \boldh_K\right]^T \in \mathbb{C}^{K \times N}$, which collects the $K$ individual RIS-UE channel vectors as its rows. Similarly, its \ac{SVD} is given by
\begin{equation}\label{eq:H_svd}
\boldH \overset{\text{SVD}}{=} \boldP \boldsymbol{\Sigma} \boldQ^H = \sum_{i=1}^{M^{\prime}} \rho_i \boldp_i \boldq_i^H,
\end{equation}
where $\boldsymbol{\Sigma} = \diag(\rho_1, \rho_2, \ldots, \rho_{M^{\prime}}) \in \mathbb{C}^{K \times N}$ is the diagonal matrix containing the singular values $\rho_i(\boldH) \in \mathbb{R}_0^+$, with $M^{\prime} = \min(K, N)$, and $i \in \mathcal{M}^{\prime} = \{1, 2,\ldots, M^{\prime}\}$. The matrices $\boldP \in \mathbb{C}^{K \times K}$ and $\boldQ \in \mathbb{C}^{N \times N}$ are unitary, where $\boldP$ contains the left singular vectors (eigenvectors of $\boldH \boldH^H$), and $\boldQ$ contains the right singular vectors (eigenvectors of $\boldH^H \boldH$). The vectors $\boldp_i$ and $\boldq_i$ denote the $i$th columns of $\boldP$ and $\boldQ$, respectively.

\subsection{Multi-UE Communication}
We now consider the \ac{MU}-\ac{MISO} scenario in which the system serves $K$ distinct \acp{UE} simultaneously. All \acp{UE} are assumed to operate in a synchronized time-division duplexing (TDD) mode, sharing a common frame and symbol clock to ensure coherent reception. Channel reciprocity is assumed for the downlink transmission, and each single-antenna \ac{UE} operates in a non-cooperative manner. Specifically, we consider the generation of $K$ uncorrelated symbols $\boldx =[ x_1, \dots, x_k, \dots, x_{K}]^T \in\mathbb{C}^{K \times 1}$, each representing an independent data stream for a specific \ac{UE}, i.e.,  $\mathbb{E}[\boldx \boldx^H] = \boldI_{K}$. By defining $\boldB = [ \boldb_1, \dots, \boldb_k, \dots, \boldb_{K}] \in\mathbb{C}^{\Nt \times K}$ as the beamforming matrix employed at the \ac{AMAF} for precoding at the feeder-side, the overall transmit power constraint is thus given by $\tr\left( \boldB \boldB^H \right) = \sum_{k=1}^K P_k =\Pt$, with $ P_k $ being the transmit power fraction allocated to the $ k $th \ac{UE} and $\Pt$ the total available transmit power.

To enable tractable analytical modeling and align with practical hardware constraints, we henceforth focus on the case in which the \ac{T-RIS} operates with a diagonal transmission matrix, i.e., $ \boldsymbol{\Phi} = \diag\left( \boldsymbol{\varphi}\right) = \diag \left(\varphi_1, \ldots, \varphi_n, \ldots, \varphi_N \right) $, where each \ac{T-RIS} element applies only a phase shift $\varphi_{n} = e^{\jmath \alpha_{n}}$ to the illumination signal emitted by the \ac{AMAF}. This configuration reflects the most hardware-efficient and scalable implementation, as it minimizes inter-element coupling and avoids the need for amplitude control. As such, it constitutes the core focus of our analysis and the baseline upon which all analytical derivations are developed. More advanced \ac{T-RIS} models featuring non-diagonal $\boldsymbol{\Phi}$ matrices, allowing inter-element interactions and full-rank transformations, are instead considered separately in Sec.~\ref{sec:non_diag} as relaxed benchmarks, used exclusively to quantify the performance gap with respect to practical and scalable solutions.

According to \cite{tse2005fundamentals}, the achievable sum-rate, expressed in bits/s/Hz, for the \ac{MU} system comprising a diagonal \ac{T-RIS} can be defined as
\begin{equation}\label{eq:sumrate}  
\Gamma = \sum_{k=1}^K \gamma_k = \sum_{k=1}^K \log_2 \left( 1 + \frac{|\bar{\boldh}_k \boldb_k|^2}{\sigmaN^2 +  \sum_{\substack{i=1 \\ i\neq k}}^{K}  |\bar{\boldh}_k \boldb_i|^2} \right),  
\end{equation}  
where $\gamma_k$ is the achievable rate of the $k$th \ac{UE}, while the denominator term $\left(\sigmaN^2 +  \sum_{\substack{i=1 \\ i\neq k}}^{K}  |\bar{\boldh}_k \boldb_i|^2\right)$ represents the interference-plus-noise power at the $k$th \ac{UE}. Consequently, the following sum-rate maximization problem can be formulated
\begin{align}\label{eq:optimizprob}
& \underset{\boldB,\, \boldsymbol{\varphi}}{\operatorname{maximize}} \quad \Gamma \\
&\text{subject to}\quad \; \boldsymbol{\Phi} = \diag\left( \boldsymbol{\varphi}\right)\, \text{,} \label{eq:diag_constr}\\
& \quad\quad\quad\quad\quad \; |\varphi_i|^2 = 1, \quad i = 1,\ldots,N  \, \text{,} \label{eq:ampl_constr}\\
& \quad\quad\quad\quad\quad \; \tr\left( \boldB \boldB^H \right) = \Pt \, \text{.} 
\end{align}
The goal of this optimization problem is to jointly design the beamforming matrix $\boldB$ at the \ac{AMAF} and the \ac{T-RIS} configuration to maximize the achievable sum rate of the \ac{MU}-\ac{MISO} system. Due to the non-convex dependence of $\Gamma$ on both the \ac{AMAF} beamforming matrix $\boldB$ and the \ac{T-RIS} coefficients vector $\boldsymbol{\varphi}$, obtaining a global solution is computationally intractable. Iterative algorithms, such as the one proposed in~\cite{choi2024wmmse}, can be employed to obtain locally optimal solutions within the feasible set. However, these methods often involve high computational complexity, limiting their practical applicability. To address this challenge, we propose a closed-form, low-complexity engineered solution for jointly designing $\boldB$ and $\boldsymbol{\varphi}$, specifically tailored to the proposed \ac{AT-RIS} antenna architecture and the considered \ac{MU}-\ac{MISO} system setup. This design leverages the underlying physics and structure of the problem to provide efficient and interpretable \ac{T-RIS} configurations with minimal computational overhead. Furthermore, the proposed solution could serve as an effective initialization point for more sophisticated optimization routines. In particular, we adopt the alternating optimization framework from~\cite{choi2024wmmse} exclusively as a refinement stage, aiming to improve the proposed engineered solution by local optimization. 

In the following, we detail the configuration strategies employed to instantiate the proposed closed-form approach under various design criteria, to maximize the \ac{MU} sum-rate while accounting for both hardware constraints and channel characteristics.

\section{AT-RIS Configuration Strategies for Sum-Rate Maximization}\label{sec:ConfigStrategies}

In this section, we develop and characterize several configuration strategies for the \ac{AT-RIS} architecture aimed at maximizing the sum-rate in the considered \ac{MU} \ac{MISO} setting. The proposed approach is structured in two stages: first, we design the beamforming matrix $\boldB$ at the \ac{AMAF} to exploit the spatial \ac{DoF} offered by the feeder-\ac{T-RIS} channel; then, we define the transmission matrix $\boldsymbol{\Phi}$ of the \ac{T-RIS} according to various design criteria.
We distinguish between two main classes of \ac{T-RIS} configurations, reflecting different trade-offs between implementation complexity and performance:
\begin{itemize}
    \item \textit{Practical low-complexity designs:} these configurations enforce a diagonal, phase-only transmission matrix, compatible with low-cost and minimally reconfigurable metasurfaces. Within this class, we propose two engineered strategies: one based on near-field focusing and one based on a phase-only \ac{MMSE} design. Both approaches aim to provide scalable and efficient support for \ac{MU} transmission while maintaining low implementation overhead. In addition, we include a \ac{PEB}-inspired design tailored to a single \ac{AT-RIS} module for a fair and comprehensive comparison with prior modular approaches.
    \item \textit{Advanced flexible designs:} in this case, the \ac{T-RIS} transmission matrix is non-diagonal, enabling arbitrary linear transformations across the surface, including amplitude modulation and element-wise coupling. Such designs require more sophisticated and costly hardware with high reconfigurability but may offer enhanced performance. We consider two representative examples within this class: one based on eigenmode alignment between the end-to-end channel components, and one derived from a \ac{MMSE} design. These non-diagonal configurations are adopted solely as performance benchmarks to assess the effectiveness of the proposed low-complexity, diagonal \ac{T-RIS} solutions.
\end{itemize}
Remarkably, this structured methodology enables a fair comparison across different design strategies, allowing us to evaluate how the engineered configuration performs under various waveform shaping criteria at the \ac{T-RIS} side, without relying on computationally intensive global optimization algorithms.

\subsection{AMAF Beamforming Configuration}

First, let us focus on the \ac{AMAF} beamforming matrix configuration. It has been shown in \cite{bartoli2023spatial} for a \ac{RIS}-aided \ac{SU}-\ac{MIMO} scenario that, when both the transmitter-\ac{RIS} and \ac{RIS}-\ac{UE} links operate in the radiative near-field region, the optimal precoding strategy for \ac{MI} maximization involves aligning the transmit beamformer with the right singular vectors of the transmitter-\ac{RIS} channel matrix. Accordingly, we define the \ac{AMAF} beamforming matrix as
\begin{equation}\label{eq:beamformersAMAF}
\boldB =  \left[\sqrt{P_1} \, \boldv_1, \dots,\sqrt{P_k} \, \boldv_k, \dots, \sqrt{P_K}\boldv_K\right] \, ,
\end{equation}
where $ \boldv_k $ is given as per \eqref{eq:G_svd}, which corresponds to assigning to each \ac{UE} an orthogonal right singular vector of $ \boldG $. After selecting the beamforming directions, we consider the associated power allocation. A straightforward strategy consists of uniformly distributing the total transmit power $ \Pt $ among the $ K $ \acp{UE}, i.e., setting $ P_k = \Pt / K $. However, improved performance might be achieved by optimizing the power distribution using the well-known waterfilling algorithm~\cite{Proakis}. 
Typically, this requires characterizing the end-to-end channel $\mathbf{H}_{\text{E}} = \mathbf{H} \boldsymbol{\Phi} \mathbf{G}$. Its \ac{SVD} is given by $\mathbf{H}_{\text{E}} \overset{\text{SVD}}{=} \mathbf{U}_{\text{E}} \boldsymbol{\Lambda}_{\text{E}} \mathbf{V}_{\text{E}}^H$, where $\boldsymbol{\Lambda}_{\text{E}} = \text{diag}(\lambda_{\text{E},1}, \dots, \lambda_{\text{E},K})$ contains the singular values and $\mathbf{V}_{\text{E}}$ is the unitary matrix of the right singular vectors. While a conventional waterfilling precoder would be constructed using these right singular vectors $\mathbf{V}_{\text{E}}$, our design intentionally retains the beamformers derived from the singular vectors of the feeder-\ac{T-RIS} channel $\mathbf{G}$ alone, as defined in~\eqref{eq:beamformersAMAF}. This choice is motivated by two key practical considerations. First, $\mathbf{G}$ is fully known at design time, as it is determined by the fixed geometry of the AMAF and the T-RIS. Second, relying on $\mathbf{V}_{\text{E}}$ would necessitate frequent pilot-based estimation of the time-varying end-to-end channel, resulting in significant signaling overhead due to the dependency of $\mathbf{H}$ on the dynamically reconfigurable \ac{T-RIS}. To provide a performance benchmark, we nonetheless use the singular values $\lambda_{\text{E},k}$ of the end-to-end channel to compute the power allocation, which prioritizes spatial streams with stronger effective gains. Although this requires full knowledge of $\mathbf{H}_{\text{E}}$, it allows us to quantify the ideal performance under perfect \ac{CSI}. The power allocated to each stream is thus given by
\begin{equation}\label{eq:waterfilling}
    P_k = \left(\mu - \frac{\sigma_{N}^2}{\lambda_{\text{E},k}^2}\right)^+, \quad k \in \mathcal{K},
\end{equation}
where $\mu$ is the water level chosen to satisfy the total power constraint $\sum_{k=1}^K P_k = P_{T}$.
Notably, this approach is consistent with the theoretical findings in \cite{bartoli2023spatial}, which demonstrate that, in the \ac{SU}-\ac{MIMO} setting with both links operating in the near-field, aligning the transmit beamformer with the singular vectors of $\mathbf{G}$ is optimal when employing a fully connected (i.e., non-diagonal) \ac{T-RIS}. Although our design constrains the $\boldsymbol{\Phi}$ matrix to a diagonal structure for practical purposes, system performance might still benefit from a non-uniform power allocation strategy, which is particularly relevant in the \ac{MU} case where \acp{UE} may experience heterogeneous path losses.

\subsection{Diagonal T-RIS Configuration}

We now shift the focus to the design of the \ac{T-RIS} transmission matrix $\boldsymbol{\Phi}$. While the work in \cite{bartoli2023spatial}, demonstrates that, in a \ac{SU}-\ac{MIMO} setting, maximizing the achievable rate can benefit from a fully connected \ac{RIS}, i.e., modeled via a non-diagonal matrix capable of implementing arbitrary linear transformations, such architectures are highly complex to realize in practice. Specifically, non-diagonal implementations require the joint control of $N^2$ coefficients and the ability to manage strong inter-element coupling across the metasurface, which significantly increases hardware complexity and cost.
To enable scalable and practical deployment, we constrain the \ac{T-RIS} to adopt a diagonal structure, allowing for the independent adjustment of only $N$ phase shifts. This simplification considerably reduces both hardware complexity and control overhead, while still offering sufficient \ac{DoF} to support spatial multiplexing at \ac{EM} level for \ac{MU} transmissions, as confirmed by our numerical study.

Inspired by the design rationale of the full-rank solution in~\cite{bartoli2023spatial}, we propose an engineered diagonal configuration that retains its core spatial alignment logic. Specifically, we define the diagonal \ac{T-RIS} phase shifts matrix as
\begin{equation}\label{eq:Phi_engineered_new}
    \boldsymbol{\Phi} = \text{diag}\left( e^{j \cdot \angle\left( \sum_{k=1}^K \boldsymbol{\psi}_k \right)} \right),
\end{equation}
where $\boldsymbol{\psi}_k \in \mathbb{C}^{N \times 1},\,k \in \mathcal{K},$ denotes a \ac{UE}-specific complex vector encapsulating the desired spatial contribution for the $k$th \ac{UE} to the overall phase profile. The set $\left\{\boldsymbol{\psi}_k\right\}_{k \in \mathcal{K}}$ is designed to shape the overall radiation pattern of the \ac{T-RIS}, thus enabling both energy focusing and inter-\ac{UE} interference mitigation. Notably, the formulation in \eqref{eq:Phi_engineered_new} maintains the phase-only constraint by applying the argument (i.e., phase) of the aggregate contribution $\sum_{k=1}^K \boldsymbol{\psi}_k$ to each element of the surface. The resulting configuration promotes constructive interference in the directions specified by the $\boldsymbol{\psi}_k$ vectors, effectively merging the individual per-\ac{UE} objectives into a unified phase design. The versatility of this approach lies in the fact that the specific structure of each $\boldsymbol{\psi}_k$ can be tailored to meet different system-level goals, such as spatial focusing, \ac{UE} separation, or channel orthogonalization. In the following, we detail three representative design strategies for generating the $\boldsymbol{\psi}_k$ vectors, each corresponding to a distinct beamforming philosophy.

\subsubsection{Focusing-Based Design}\label{sec:focusing}

The first strategy constructs the $\boldsymbol{\psi}_k$ vectors by combining spatial mode selection with near-field beam focusing capabilities, in line with the theoretical structure of the optimal non-diagonal solution \cite{torcolacci2023oam}. Specifically, we define
\begin{align}\label{eq:psi_foc}
   \boldsymbol{\psi}_k = \boldu_k^H \odot \boldf_k^{\text{(NF)}} \, , \quad k \in \mathcal{K},
\end{align}
where $\boldu_k^H$ denotes the $k$th row of the matrix $\boldU$ defined in~\eqref{eq:G_svd}, and $\boldf_k^{\text{(NF)}} \in \mathbb{C}^{N \times 1}$ represents the near-field focusing vector pointing towards the $k$th \ac{UE} given by
\begin{equation}\label{eq:f_k_nf}
    \boldf_k^{\text{(NF)}} =  \left[e^{j \kappa d_{1}^{(k)}}, \dots, e^{j \kappa d_{i}^{(k)}}, \dots, e^{j \kappa d_{N}^{(k)}}\right] \in \mathbb{C}^{N \times 1}
\end{equation}
where $\kappa = 2\pi/\lambda$ is the wavenumber, and $d_i^{(k)} = \left\| \boldp_{\text{RIS}, i} - \boldp_{\text{R}}^{(k)} \right\|$ denotes the Euclidean distance between the $i$th \ac{T-RIS} element and the $k$th \ac{UE}. The vector $\boldf_k^{\text{(NF)}}$ implements a phase compensation corresponding to the conjugate of the near-field channel response in \eqref{eq:hk_coeff}, thus enabling coherent energy focusing at the intended \ac{UE} location. By jointly combining near-field beam focusing and alignment with the spatial modes of the feeder channel, this approach enhances signal discrimination across \acp{UE} while satisfying the phase-only constraint of the diagonal \ac{T-RIS}.
Notably, as the distance between the \ac{T-RIS} and the \ac{UE} increases, $\boldf_k^{\text{(NF)}}$ gradually converges to a classical far-field steering vector, thereby ensuring continuity and robustness of the design across both near- and far-field propagation regimes.

\subsubsection{MMSE-Based Design}\label{sec:mmse_diag}

As an alternative to near-field focusing, the $\boldsymbol{\psi}_k$ vectors can be derived using classical \ac{MIMO} beamforming techniques, such as the \ac{MMSE} precoder \cite{Biglieri_book}. Specifically, the \ac{MMSE} precoding matrix is given by
\begin{equation}\label{eq:ell_mmse}
    \boldL = \eta \left( \sigma_{\text{TX}}^2 \boldI_N + \boldH^H \boldH \right)^{-1} \boldH^H \in \mathbb{C}^{N \times K},
\end{equation}
where $\sigma_{\text{TX}}^2 = \delta_{\text{TX}} \left( \|\boldH\|_{\mathrm{F}}^2/N \right)$ is a regularization term controlled by the scalar hyperparameter $\delta_{\text{TX}} > 0$, and $\eta = 1 / \|\boldL\|_{\mathrm{F}}$ is a normalization factor enforcing the transmit power constraint. This guarantees that the \ac{T-RIS}, being a passive device, does not amplify or attenuate the impinging signals. Letting $\boldsymbol{\ell}_k$ denote the $k$th column of $\boldL$, the corresponding \ac{T-RIS} transmission vector is constructed as
\begin{equation}\label{eq:psi_mmse_diag}
    \boldsymbol{\psi}_k = \boldu_k^H \odot \boldsymbol{\ell}_k \,, \quad k \in \mathcal{K},
\end{equation}
thereby combining the eigenstructure of the feeder-\ac{T-RIS} wireless channel with the interference-aware digital precoding at the \ac{T-RIS} defined by the \ac{MMSE} strategy. As in the focusing-based case, the final diagonal \ac{T-RIS} matrix is obtained from~\eqref{eq:Phi_engineered_new}, ensuring phase-only operation while fostering constructive signal combining at the receiving \acp{UE}. Notably, this \ac{MMSE}-inspired design is particularly well-suited to scenarios characterized by high \ac{UE} density or strong inter-\ac{UE} channel correlation, where focusing alone may fail to ensure sufficient interference suppression. By incorporating \ac{MU} interference mitigation directly into the \ac{T-RIS} beamforming design, this approach achieves a more effective allocation of spatial and power resources. It thus offers a practical compromise between digital beamforming capabilities and \ac{EM}-level waves control in large-scale \ac{MU} deployments.

\subsubsection{PEB-Based Design with T-RIS Partitioning}\label{sec:peb}

Another representative strategy, originally proposed in~\cite{TiwCai2:C23} for \ac{SU} far-field communication, is the so-called \ac{PEB} design. This approach constrains the \ac{AMAF} beamforming vector to align with the principal right-singular vector of the feeder-\ac{T-RIS} channel $\boldG$, i.e., $\boldb_1 = \boldv_1$. The corrisponding \ac{T-RIS} transmission vector for the target \ac{UE} (denoted here as the generic $k$th \ac{UE}) is then given by
\begin{equation}\label{eq:peb_caire}
\boldsymbol{\psi}_k = \boldu_1^H \odot \boldf_k^{\text{(FF)}} \, \, , \, k \in \mathcal{K} ,
\end{equation}
where
\begin{equation}\label{eq:f_k_ff}
   \boldf_k^{\text{(FF)}} = \left[ e^{j \kappa \, \mathbf{n}(\boldsymbol{\Theta}_k)^T \boldp_{\text{RIS}, j}} \right]_{j=1}^{N} \in \mathbb{C}^{N \times 1}
\end{equation}
is the far-field steering vector towards the intended \ac{UE} and $\boldn( \boldsymbol{\Theta}_k) = \left[ \sin(\theta_k) \cos(\phi_k) , \sin(\theta_k) \sin(\phi_k) ,\cos(\theta_k) \right]^T $ is the unit-norm direction vector specifying the departure angle under the adopted angular convention. The final \ac{T-RIS} phase shift matrix is then computed as $\boldsymbol{\Phi} = \diag\left(\boldsymbol{\psi}_k  \right)$. It is worth noting that, unlike the near-field focusing vector in~\eqref{eq:f_k_nf}, which accounts for the exact distance between each \ac{T-RIS} element and the target \ac{UE}, the far-field steering vector in~\eqref{eq:f_k_ff} relies solely on the direction of departure $\boldsymbol{\Theta}_k$, thus assuming planar \ac{EM} wavefronts. This approximation is valid when the \ac{UE} lies in the far-field region of the \ac{T-RIS}, such that path length variations across the array can be neglected. While straightforward and effective for \ac{SU} transmission, the approach in~\cite{TiwCai2:C23} extends to the \ac{MU} case by stacking $K$ identical and equispaced \ac{AT-RIS} modules, each exclusively assigned to a specific \ac{UE} and independently implementing the \ac{SU} beamforming scheme \eqref{eq:peb_caire} proposed therein. However, this architectural solution becomes increasingly impractical in dense \ac{UE} scenarios due to the larger spatial footprint and hardware redundancy associated with deploying multiple physically separated \ac{AT-RIS} modules. Moreover, such an architecture is inherently tailored to a fixed number of \acp{UE}, limiting its adaptability in dynamic environments. Any variation in the actual \ac{UE} count may lead to under-utilization or saturation of the available \ac{T-RIS} modules, thereby reducing system efficiency and scalability.

In this regard, to enable a fair benchmarking between their modular architecture comprising several disjoint units, and our antenna solution based on a single contiguous \ac{T-RIS} surface, we emulate a similar operation by partitioning the unified \ac{T-RIS} into $K$ disjoint sectors, each assigned to a given \ac{UE}. Accordingly, we define a set of \ac{UE}-specific feeder-\ac{T-RIS} channel matrices $\left\{\boldG_k\right\}_{k \in \mathcal{K}}  \in \mathbb{C}^{N \times \Nt}$, where each $\boldG_k$ retains only the entries of the global channel matrix $\boldG$ corresponding to the subset of \ac{T-RIS} elements assigned to the $k$th \ac{UE}, while all other entries are zeroed. Formally, let $\mathcal{S}_k \subseteq \mathcal{N} $ denote the index set of \ac{T-RIS} elements associated with the $k$th sector. Then, the entries of $\boldG_k$ are defined as
\begin{equation}
\left[\boldG_k\right]_{j,i} =
\begin{cases}
g_{j,i}, & \text{if } j \in \mathcal{S}_k, \\
0, & \text{otherwise}
\end{cases}
\quad \forall i \in \mathcal{N}_{\text{T}}, \quad j \in \mathcal{N}\,.
\end{equation}
As a result, each matrix $\boldG_k$ thus isolates the contribution of the $k$th sector by masking all entries outside $\mathcal{S}_k$, effectively modeling a virtual partition of the \ac{T-RIS} for per-\ac{UE} beamforming. Based on this sector-specific channel, for each \ac{UE} $k \in \mathcal{K}$ we compute the \ac{SVD} of $\boldG_k$ as
\begin{equation}
\boldG_k \overset{\text{SVD}}{=} \boldU_k \boldsymbol{\Sigma}_k \boldV_k^H = \sum_{i=1}^{M^{\prime\prime}} \beta_i^{(k)} \boldu_i^{(k)} \left(\boldv_i^{(k)}\right)^H \,,
\end{equation}
where $\boldU_k \in \mathbb{C}^{N \times N}$ and $\boldV_k \in \mathbb{C}^{\Nt \times \Nt}$ are unitary matrices containing the left and right singular vectors, respectively, and $\boldsymbol{\Sigma}_k = \operatorname{diag}(\beta_1^{(k)}, \beta_2^{(k)}, \ldots, \beta_{M^{\prime\prime}}^{(k)}) \in \mathbb{C}^{N \times \Nt}$ is the diagonal matrix of singular values, with $M^{\prime\prime} = \min(\Nt, N)$, and $i \in \mathcal{M}^{\prime\prime} = \{1,2, \ldots, M^{\prime\prime}\}$. In line with the method proposed in \cite{TiwCai2:C23}, the beamforming vector for the $k$th \ac{UE} is then selected as $\boldb_k = \boldv_1^{(k)} \in \mathbb{C}^{\Nt \times 1}$, i.e., the first right singular vector of $\boldG_k$, corresponding to its strongest spatial mode. To compute the \ac{T-RIS} transmission vector for \ac{UE} $k$, we then follow the same structure used in the focusing-based design, but replace the spatial mode selection vector with the conjugate of the first left singular vector of $\boldG_k$, i.e.,
\begin{equation}\label{eq:psi_peb_diag}
\boldsymbol{\psi}_k = \left( \boldu_1^{(k)} \right)^H \odot   \boldf_k^{\text{(NF)}} \,,
\end{equation}
where $  \boldf_k^{\text{(NF)}} $ is the near-field focusing vector as per \eqref{eq:f_k_nf}, which inherently reduces to the far-field steering vector~\eqref{eq:f_k_ff} when the \ac{UE} lies in the far-field region. Finally, the global \ac{T-RIS} phase shift matrix is constructed by aggregating the $K$ per-\ac{UE} vectors in \eqref{eq:psi_peb_diag} according to~\eqref{eq:Phi_engineered_new}. Importantly, while $\boldf_k^{\text{(NF)}}$ includes phase terms for all \ac{T-RIS} elements, the structure of $\boldu_1^{(k)}$ ensures that only the elements in $\mathcal{S}_k$ contribute non-trivially to the product. This guarantees that the beamforming operation remains confined to the sector assigned to \ac{UE} $k$, thus preserving the intended sectorization. Therefore, the sum $\sum_{k=1}^K \boldsymbol{\psi}_k$ in~\eqref{eq:Phi_engineered_new} results in a non-overlapping aggregation of per-\ac{UE} contributions, effectively preserving sector isolation while allowing for a compact and unified phase shift matrix expression. As a result, this method effectively emulates the modular beamforming strategy proposed in \cite{TiwCai2:C23} using a single monolithic \ac{AT-RIS} module, while enabling per-\ac{UE} beam design and sector-specific operation. In addition, this approach inherently captures potential spillover and interference effects between adjacent \ac{T-RIS} sectors, thereby providing a more realistic model of practical \ac{T-RIS} implementations. 

\subsection{Non-diagonal T-RIS Configuration}\label{sec:non_diag}

To enable a meaningful performance comparison and quantify the relative loss incurred by adopting scalable, hardware-efficient designs, we consider a generalized \ac{T-RIS} model in which both the diagonal and the phase-only constraints in \eqref{eq:diag_constr}-\eqref{eq:ampl_constr} are relaxed. Specifically, we allow the transmission matrix $\boldsymbol{\Phi}$ to assume a fully non-diagonal structure with arbitrary complex-valued entries, enabling joint amplitude and phase modulation across all \ac{T-RIS} elements. Although such a configuration is not aligned with the practical and scalable \ac{AT-RIS} architecture promoted in this work, which relies on a diagonal \ac{T-RIS}, it provides a valuable reference to evaluate how closely practical solutions can approach the performance of more complex and less cost-effective alternatives. In this context, we introduce two representative non-diagonal \ac{T-RIS} configurations, differing in their beamforming strategies and receiver assumptions, and use them as upper-bound benchmarks in the numerical analysis.

\subsubsection{Eigenmodes-Based Design} \label{sec:eig_nd}

This configuration employs the optimal non-diagonal \ac{T-RIS} scheme originally derived in~\cite{bartoli2023spatial} for maximizing \ac{MI} in \ac{SU} near-field systems, based on spatial eigenmode multiplexing. In this approach, the \ac{T-RIS} performs a full-rank linear transformation that aligns the eigenmodes of the feeder-\ac{T-RIS} and \ac{T-RIS}-receiver channels. The optimal transformation matrix is given by
\begin{equation}\label{eq:Psi_eigen}
\boldsymbol{\Phi} = \boldQ \, \boldU^H,
\end{equation}
where $\boldU \in \mathbb{C}^{N \times N}$ contains the left singular vectors of the feeder-\ac{T-RIS} channel $\boldG$ (as defined in~\eqref{eq:G_svd}), and $\boldQ \in \mathbb{C}^{N \times N}$ includes the right singular vectors of the receiver-side channel matrix $\boldH$ (as per~\eqref{eq:H_svd}). We note that this optimal configuration not only exploits the spatial eigenmode alignment enabled by a non-diagonal \ac{T-RIS} transformation, but also relies on a transmit-side power allocation following the classical waterfilling profile over the composite singular values of $\boldG$ and $\boldH$. Accordingly, the \ac{AMAF} beamforming vectors must be weighted based on the waterfilling solution given in~\eqref{eq:waterfilling} to approach the optimal spectral efficiency predicted in~\cite{bartoli2023spatial}. In particular, selecting the \ac{T-RIS} matrix as in~\eqref{eq:Psi_eigen} diagonalizes the end-to-end channel $\mathbf{H}_{\text{E}}$, whose singular values reduce to the element-wise product $\lambda_{E,k} = \xi_k \rho_k$, where $\xi_k$ and $\rho_k$ denote the $k$th singular values of the feeder-\ac{T-RIS} channel $\mathbf{G}$ and the \ac{T-RIS}-\acp{UE} channel $\mathbf{H}$, respectively. Power allocation across streams then follows the waterfilling solution in~\eqref{eq:waterfilling}, applied to these composite gains
\begin{equation}\label{eq:waterfill_nondiag}
    P_k = \left( \mu - \frac{\sigma_N^2}{\xi_k^2 \rho_k^2} \right)^+, \quad k \in \mathcal{K}.
\end{equation}
This ensures optimal power distribution over the orthogonal eigenmodes induced by the joint channel structure. Moreover, the original design assumes a multi-antenna receiver capable of coherent combining, which does not directly extend to conventional \ac{MU} systems with independent single-antenna \acp{UE}. Therefore, to emulate the original setup and obtain an upper bound for our scenario, we model the receiver as a cooperative entity performing optimal linear combining, thus ideally assuming inter-\ac{UE} cooperation. Specifically, we define a global receive filter matrix $\boldP = [\boldp_1, \ldots, \boldp_K]$, where each $\boldp_k$ is the $k$th column of $\boldP$ as derived from the \ac{SVD} of $\boldH$ in~\eqref{eq:H_svd}. Under this ideal cooperative framework, the joint design of the AMAF precoder $\boldB$, the T-RIS transformation $\boldsymbol{\Phi}$, and the receive filter $\boldP$ effectively diagonalizes the end-to-end channel. This process steers each data stream along a distinct orthogonal eigenmode of the end-to-end channel, thus completely nullifying inter-\ac{UE} interference after receive processing. Consequently, the achievable sum-rate is determined by the \ac{SNR} of each independent spatial stream, leading to
\begin{equation}
    \Gamma = \sum_{k=1}^K \log_2 \left( 1 + \frac{\left|\mathbf{p}_k^H \bar{\mathbf{h}}_k \mathbf{b}_k\right|^2}{\sigma_N^2} \right) \,,
\end{equation}
which assumes full \ac{CSI} and centralized joint processing at the receiver side.

\subsubsection{MMSE-based Design} \label{sec:mmse_nd}

In this variant, the non-diagonal \ac{T-RIS} is designed to incorporate the classical \ac{MMSE} precoder directly into the \ac{T-RIS} response. Specifically, we define
\begin{equation}\label{eq:MMSEnondiag}
\boldsymbol{\Phi} = \boldL \, \boldU^H,
\end{equation}
where $\boldL \in \mathbb{C}^{N \times K}$ is the \ac{MMSE} precoding matrix as defined in~\eqref{eq:ell_mmse}. Unlike the eigenmode-based approach, this formulation assumes no cooperation among \acp{UE}, and the sum-rate is computed under independent per-\ac{UE} decoding, without linear combining at the receiver, i.e., as in~\eqref{eq:sumrate}. This makes it a relevant benchmark for \ac{MU} scenarios with non-cooperative \acp{UE}, consistent with practical deployments where joint decoding is unfeasible. In contrast to the eigenmode-based design, the solution in~\eqref{eq:MMSEnondiag} inherently performs power allocation at the \ac{T-RIS}, balancing inter-\ac{UE} interference and noise amplification via amplitude modulation at the \ac{T-RIS}. This enables the \ac{AMAF} to operate with uniform power allocation across \acp{UE}, delegating the fine-grained per-\ac{UE} power distribution and interference management to the metasurface. In this way, part of the beamforming complexity is offloaded from the feeder to the \ac{T-RIS}, introducing architectural decoupling that may simplify implementation in large-scale systems. Taken together, these features make the \ac{MMSE}-based transformation a practically motivated and scalable alternative within the class of non-diagonal \ac{T-RIS} schemes. While not theoretically optimal, it serves as an intermediate benchmark between highly idealized cooperative solutions and the diagonal architectures promoted in this work. 

Building on this, both non-diagonal configurations effectively establish performance upper bounds for cooperative and non-cooperative \acp{UE}, respectively, as they bypass the hardware constraints inherent to diagonal, phase-only \ac{T-RIS} implementations. By enabling full-rank transformations across the surface, these schemes provide valuable references to gauge how closely practical, scalable architectures can approach theoretical limits.

To systematically evaluate and compare all considered \ac{T-RIS} configurations, both diagonal and non-diagonal, we introduce Jain’s fairness index as a standard metric to quantify the equity in rate distribution among \acp{UE}, defined as \cite{jain1984quantitative}
\begin{equation}
\mathcal{J}(\{\gamma_k\}) = \frac{\left(\sum_{i=1}^K \gamma_i \right)^2}{K \sum_{i=1}^K \gamma_i^2}\, \in\, [0,1] \,,\; k \in \mathcal{K}\,.
\end{equation}
Unlike average rate or rate variance, which capture only central tendency or dispersion, respectively, the Jain index offers a normalized, bounded metric that reflects the equitability of rate allocation across \acp{UE}, regardless of the absolute rate values. This is particularly relevant in \ac{MU} deployments, where high sum-rate values may mask substantial disparities in individual \ac{UE} performance, ultimately degrading quality of service and \ac{UE} experience.

\section{Numerical Results}\label{sec:NumericalResults}

\subsection{Simulation Setup}

In our numerical study, \ac{MU} transmission is performed over a narrow frequency band $\Delta f=120\,\mathrm{MHz}$ centered at $f_{\text{c}}=28\, \mathrm{GHz}$, corresponding to a wavelength of $\lambda\simeq 0.01\, \mathrm{m}$. This setting can be associated with the adoption of a single subcarrier or resource block in an \ac{OFDM} system, suitable for communication and sensing purposes. The noise power spectral density is set to $\sigmaN^2 = -170\,\mathrm{dBm/Hz}$, while the transmit power is fixed to $\Pt = 10\,\mathrm{mW}$. The \ac{AMAF} array consists of $\Nt = 16$ active antenna elements arranged as a $(\lambda/2)$-spaced square \ac{UPA} on the $(x,z)$ plane. The \ac{T-RIS} is positioned in a paraxial configuration with respect to the AMAF, with its center located at a distance of $8\lambda$ along the $y$-axis from the \ac{AMAF}. It has a square aperture of size $(25\lambda \times 25\lambda)\,\mathrm{m}^2$, with meta-atoms that are spaced $\lambda/2$ apart, resulting in a total of $N = 2500$ elements. The overall \ac{AT-RIS} structure is deployed on the same horizontal plane as the \acp{UE}, with zero elevation and no tilt. Given this geometry, the Fraunhofer distance is located at approximately $d_{\text{FF}} \approx 27 \, \mathrm{m}$ along the $y$ direction, hence discriminating among the radiative near-field and far field regions.
Moreover, each \ac{UE} is equipped with a single receiving antenna. The total number and spatial distribution of \acp{UE} vary depending on the deployment scenario considered in the numerical analysis, ranging from few-\ac{UE} cases (e.g., $K=2$) to scenarios with a larger number of \acp{UE} ($2 < K \leq 16$), as further detailed in the following subsections.
In addition, the directional antenna gain functions $G_{\text{T}}(\boldsymbol{\Theta})$ and $G_{\text{R}}(\boldsymbol{\Theta})$, respectively associated with the \ac{AMAF} and \acp{UE} antenna elements, are assumed to be identical for all array elements and are given by $ G_{\text{T}}(\boldsymbol{\Theta}) = G_{\text{R}}(\boldsymbol{\Theta}) = 2 \sin(\theta) \sin(\phi)$. This follows the directional patch antenna model also adopted in~\cite{TiwCai:C24}, where the element achieves maximum radiation in the broadside direction\footnote{This simplified antenna gain model isolates the fundamental multiplexing properties of the \ac{AT-RIS} architecture, prioritizing analytical tractability over exact realism. Future work will explore the impact of practical antenna patterns on system performance.}.

In our numerical evaluation, we compare multiple \ac{AT-RIS} configuration strategies that differ in two key design aspects: (\textit{i}) the power allocation method used to construct the \ac{AMAF} beamforming matrix $\boldB$, and (\textit{ii}) the criterion employed to generate the \ac{UE}-specific vectors $\left\{ \boldsymbol{\psi}_k \right\}_{k \in \mathcal{K}}$ used to build the \ac{T-RIS} transmission matrix $\boldsymbol{\Phi}$. Each considered scheme results from a specific combination of these two components, leading to different trade-offs between complexity and performance. Specifically, we consider the following configurations:

\begin{itemize}

\item \textit{D-FOC-U}: Diagonal \ac{T-RIS} configuration employing the focusing-based approach in \eqref{eq:psi_foc} and uniform power allocation at the \ac{AMAF}.  

\item \textit{D-FOC-W}: Diagonal \ac{T-RIS} configuration utilizing the focusing-based design in \eqref{eq:psi_foc} and waterfilling-based power allocation at the \ac{AMAF} as per \eqref{eq:waterfilling}.  

\item \textit{D-MMSE-U}: Diagonal \ac{T-RIS} configuration adopting the \ac{MMSE}-based method in \eqref{eq:psi_mmse_diag}, with $\delta_{\text{TX}}  =  10^{-8}$, and uniform power allocation at the \ac{AMAF}. 

\item \textit{D-PEB-U}: Diagonal \ac{T-RIS} configuration implementing the \ac{PEB}-inspired design in \eqref{eq:psi_peb_diag} and uniform power allocation at the \ac{AMAF}.  
\item \textit{ND-EIG-W}: Non-diagonal \ac{T-RIS} configuration employing the configuration \eqref{eq:Psi_eigen} and waterfilling power allocation at the \ac{AMAF} as per \eqref{eq:waterfill_nondiag}. 

\item \textit{ND-MMSE-U}: Non-diagonal \ac{T-RIS} configuration utilizing \ac{MMSE}-based strategy in \eqref{eq:MMSEnondiag}, with $\delta_{\text{TX}}  =  10^{-8}$, and uniform power allocation at the \ac{AMAF}.

\end{itemize}

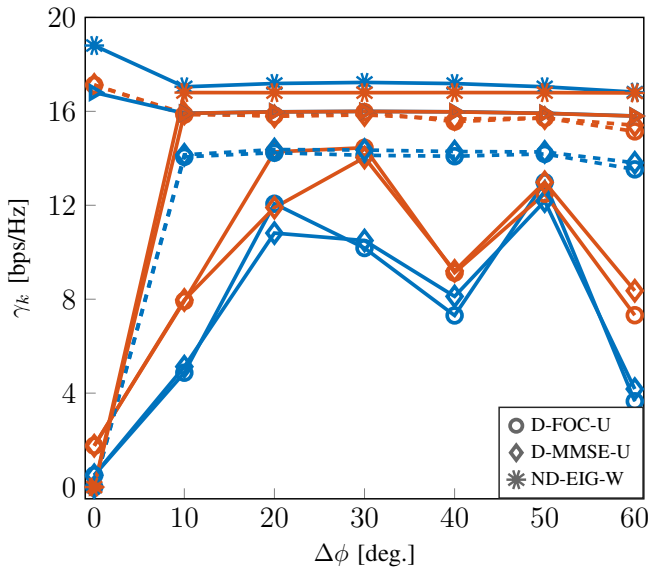
\begin{figure}[t!]
    \centering
    \hspace{-5mm}\resizebox{\linewidth}{!}{



\definecolor{mycolor1}{rgb}{0.00000,0.44700,0.74100}%
\definecolor{mycolor2}{rgb}{0.85000,0.32500,0.09800}%
\definecolor{darkgray}{rgb}{0.4,0.4,0.4}


\begin{tikzpicture}
\begin{axis}[%
    width=\columnwidth,
    width=0.9\columnwidth,
    scale only axis,
  xmin=-1,
    xmax=61,
    xtick={0,10, 20, 30, 40, 50, 60},
    ytick={0,4,8,12,16,20},
    xlabel style={font=\fontsize{12}{14}\selectfont, color=black},
    xlabel={$\Delta \phi$ [deg.]},
    ymin=-0.5,
    ymax=20,
    ylabel style={font=\fontsize{12}{14}\selectfont, color=black},
    ylabel={$\gamma_k$ [bps/Hz]},
    axis background/.style={fill=white},
  axis lines=box, 
    line width=0.6pt, 
xmajorgrids=false,
ymajorgrids=false,
    tick label style={font=\fontsize{12}{14}\selectfont},
    label style={font=\fontsize{12}{14}\selectfont},
    title style={font=\fontsize{12}{14}\selectfont},
    legend style={
    at={(1,0)},
    anchor=south east,
    legend cell align=left,
    align=left,
    draw=white!15!black,
    font=\fontsize{8}{10}\selectfont
},
axis lines=box,
xtick pos=bottom,
ytick pos=left,
xticklabel pos=lower,
yticklabel pos=left,
extra x ticks={},
extra y ticks={},
extra x tick style={draw=none},
extra y tick style={draw=none}
]

\addplot [ forget plot, color=mycolor1, line width=1.5pt,mark size= 3pt, mark=o, mark options={line width=1.5pt, solid, mycolor1}] table[row sep=crcr]{%
0	0.504647422534105\\
10	4.86985463782811\\
20	12.0652753732609\\
30	10.1795415314164\\
40	7.30726774125517\\
50	12.9806280543816\\
60	3.65322274437107\\
};

\addplot [ forget plot, color=mycolor2, line width=1.5pt,mark size= 3pt, mark=o, mark options={line width=1.5pt, solid, mycolor2}] table[row sep=crcr]{%
0	1.76027003534244\\
10	7.92327802098224\\
20	14.2741208219457\\
30	14.4498404063634\\
40	9.13038123608929\\
50	12.4920271476355\\
60	7.31737796293044\\
};

\addplot [ forget plot, color=mycolor1, dashed, line width=1.5pt,mark size= 3pt, mark=o, mark options={line width=1.5pt, solid, mycolor1}]  table[row sep=crcr]{%
0	0\\
10	14.060818765994\\
20	14.2374396743051\\
30	14.1274823327556\\
40	14.0884717949142\\
50	14.1863162153006\\
60	13.5282432222502\\
};

\addplot [ forget plot, color=mycolor2, dashed, line width=1.5pt,mark size= 3pt, mark=o, mark options={line width=1.5pt, solid, mycolor2}]  table[row sep=crcr]{%
0	17.1226397894395\\
10	15.8602892049801\\
20	15.8253966564818\\
30	15.9686950235659\\
40	15.5634202089692\\
50	15.7156712814581\\
60	15.1462967702763\\
};





\addplot [ forget plot, color=mycolor1, line width=1.5pt,mark size= 4pt, mark=diamond, mark options={line width=1.5pt, solid, mycolor1}]  table[row sep=crcr]{%
0	0.504653798673353\\
10	5.12727144985181\\
20	10.8159306418323\\
30	10.497521651571\\
40	8.11974887876286\\
50	12.1754831694201\\
60	4.17390520203955\\
};

\addplot [ forget plot, color=mycolor2, line width=1.5pt,mark size= 4pt, mark=diamond, mark options={line width=1.5pt, solid, mycolor2}]  table[row sep=crcr]{%
0	1.76025481094275\\
10	7.9674028046113\\
20	11.9010179985016\\
30	14.0323623935827\\
40	9.20466598645204\\
50	12.9798489388745\\
60	8.36015129123646\\
};

\addplot [ forget plot, color=mycolor1, dashed, line width=1.5pt,mark size= 4pt, mark=diamond, mark options={line width=1.5pt, solid, mycolor1}]  table[row sep=crcr]{%
0	0\\
10	14.1484971720272\\
20	14.3728283266404\\
30	14.3522670195741\\
40	14.2880295512896\\
50	14.2709709439193\\
60	13.8089885682356\\
};

\addplot [ forget plot, color=mycolor2, dashed, line width=1.5pt,mark size= 4pt, mark=diamond, mark options={line width=1.5pt, solid, mycolor2}]  table[row sep=crcr]{%
0	17.1226375454445\\
10	15.9151177805735\\
20	15.7890047782702\\
30	15.8465441320698\\
40	15.6748397042105\\
50	15.7167193814164\\
60	15.4065624408727\\
};

\addplot [ forget plot, color=mycolor1, line width=1.5pt,mark size= 3pt, mark=triangle, mark options={line width=1.5pt, solid, rotate=270, mycolor1}]  table[row sep=crcr]{%
0	16.803027812658\\
10	15.9185917123977\\
20	15.9796442151064\\
30	15.9990127351582\\
40	15.9805568179865\\
50	15.9184360965452\\
60	15.8028708750478\\
};

\addplot [ forget plot, color=mycolor2, line width=1.5pt,mark size= 3pt, mark=triangle, mark options={line width=1.5pt, solid, rotate=270, mycolor2}]  table[row sep=crcr]{%
0	0\\
10	15.915139864989\\
20	15.9761856736675\\
30	15.995557159153\\
40	15.9770969312251\\
50	15.9149754956385\\
60	15.7994145737115\\
};

\addplot [ forget plot, color=mycolor1, line width=1.5pt,mark size=4pt, mark=10-pointed star, mark options={line width=1pt, solid, rotate=90, mycolor1}]  table[row sep=crcr]{%
0	18.8030184573641\\
10	17.0447358995625\\
20	17.1860465031915\\
30	17.2313387616991\\
40	17.1830505946319\\
50	17.0456063307998\\
60	16.8186077451276\\
};

\addplot [ forget plot, color=mycolor2, line width=1.5pt,mark size=4pt, mark=10-pointed star,  mark options={line width=1pt, solid, rotate=90, mycolor2}]  table[row sep=crcr]{%
0	0\\
10	16.7991235036962\\
20	16.7956319403739\\
30	16.7954886093406\\
40	16.7995401152751\\
50	16.7980933167978\\
60	16.7838231838062\\
};


\addlegendimage{mark size=2.5pt, only marks, line width=1.5pt, mark=o, mark options={line width=1.5pt, solid, darkgray}} 
\addlegendentry{D-FOC-U}

\addlegendimage{mark size=3pt, only marks, line width=1.5pt, mark=diamond, mark options={line width=1.5pt, solid, darkgray}} 
\addlegendentry{D-MMSE-U}

\addlegendimage{mark size=4pt, only marks, line width=1.5pt, mark=10-pointed star,  mark options={line width=1pt, solid, rotate=90, darkgray}} 
\addlegendentry{ND-EIG-W}


\end{axis}
\end{tikzpicture}%
}  
\caption{Per-\ac{UE} achievable rate $\gamma_k$ as a function of the angular separation $\Delta\phi$ between two \acp{UE} placed at $d=10\,\mathrm{m}$ for different \ac{AT-RIS} configuration strategies. Blue and orange curves correspond to UE1 and UE2, respectively. Solid lines indicate the proposed configuration, while dashed lines denote its optimized counterpart obtained via the algorithm in~\cite{choi2024wmmse}.}
  \label{fig:1A_perUErate}
\end{figure}

\begin{figure*}[t!]
    \centering
    \resizebox{\textwidth}{!}{ 
        \begin{minipage}[t]{0.48\textwidth}
            \centering
            \hspace{-5mm}\resizebox{\linewidth}{!}{



\definecolor{mycolor1}{rgb}{0.00000,0.44700,0.74100}%
\definecolor{mycolor2}{rgb}{0.85000,0.32500,0.09800}%
\definecolor{mycolor3}{rgb}{0.92900,0.69400,0.12500}%
\definecolor{mycolor4}{rgb}{0.49400,0.18400,0.55600}%
\definecolor{mycolor5}{rgb}{0.46600,0.67400,0.18800}%
\definecolor{mycolor6}{rgb}{0.63500,0.07800,0.18400}%
\definecolor{mycolor7}{rgb}{0.30196,0.74510,0.93333}%
\definecolor{darkgray}{rgb}{0.4,0.4,0.4}
\definecolor{mygreen}{rgb}{0.0, 0.6, 0.3} 
\definecolor{mybrick}{rgb}{0.63500,0.07800,0.18400}%


\begin{tikzpicture}
\begin{axis}[%
    width=\columnwidth,
    width=0.9\columnwidth,
    scale only axis,
  xmin=-1,
    xmax=61,
    xtick={0,10, 20, 30, 40, 50, 60},
    ymin=1.5,
    ymax=35,
    ytick={5,10,15, 20, 25, 30, 35},
    xlabel style={font=\fontsize{12}{14}\selectfont, color=black},
    xlabel={$\Delta \phi$ [deg.]},
    ylabel style={font=\fontsize{12}{14}\selectfont, color=black},
    ylabel={$\Gamma$ [bps/Hz]},
    title={$d = 10 $ m},
    axis background/.style={fill=white},
  axis lines=box, 
    line width=0.6pt, 
xmajorgrids=false,
ymajorgrids=false,
    tick label style={font=\fontsize{12}{14}\selectfont},
    label style={font=\fontsize{12}{14}\selectfont},
    title style={font=\fontsize{12}{14}\selectfont},
    legend style={
    at={(1,0)},
    anchor=south east,
    legend cell align=left,
    align=left,
    draw=white!15!black,
    font=\fontsize{8}{10}\selectfont
},
axis lines=box,
xtick pos=bottom,
ytick pos=left,
xticklabel pos=lower,
yticklabel pos=left,
extra x ticks={},
extra y ticks={},
extra x tick style={draw=none},
extra y tick style={draw=none}
]

\addplot [ forget plot, color=mycolor1,line width=1.5 pt, mark size=3pt, mark=o, mark options={line width=1.5pt, solid, mycolor1}]  table[row sep=crcr]{%
0	2.26491745787655\\
10	12.7931326588103\\
20	26.3393961952065\\
30	24.6293819377798\\
40	16.4376489773445\\
50	25.4726552020171\\
60	10.9706007073015\\
};

\addplot [ forget plot, color=mycolor1, dashed,line width=1.5 pt, mark size=3pt, mark=o, mark options={line width=1.5pt, solid, mycolor1}]  table[row sep=crcr]{%
0	17.1226397894395\\
10	29.921107970974\\
20	30.0628363307868\\
30	30.0961773563215\\
40	29.6518920038834\\
50	29.9019874967587\\
60	28.6745399925264\\
};



\addplot [ forget plot, color=mycolor3,line width=1.8 pt,mark size=4pt,   mark=diamond, mark options={line width=1.5pt, solid, mycolor3}]  table[row sep=crcr]{%
0	2.2649086096161\\
10	13.0946742544631\\
20	22.7169486403339\\
30	24.5298840451538\\
40	17.3244148652149\\
50	25.1553321082946\\
60	12.534056493276\\
};

\addplot [ forget plot, color=mycolor3, dashed,line width=1.8 pt,mark size=4pt,   mark=diamond, mark options={line width=1.5pt, solid, mycolor3}]  table[row sep=crcr]{%
0	17.1226375454445\\
10	30.0636149526008\\
20	30.1618331049105\\
30	30.1988111516439\\
40	29.9628692555001\\
50	29.9876903253358\\
60	29.2155510091083\\
};

\addplot [ forget plot, color=mycolor4,line width=1.5 pt, mark size=3pt,   mark=triangle, mark options={line width=1.5pt, solid, mycolor4}]  table[row sep=crcr]{%
0	2.00014082832817\\
10	17.5583723819618\\
20	19.4548564137378\\
30	19.5925817884049\\
40	20.2308034150248\\
50	19.5880911126246\\
60	19.0712206531348\\
};

\addplot [ forget plot, color=mycolor4, dashed,line width=1.5 pt, mark size=3pt,   mark=triangle, mark options={line width=1.5pt, solid, mycolor4}] table[row sep=crcr]{%
0	17.0713946765812\\
10	30.7159818137339\\
20	30.5732252495613\\
30	30.5745846998666\\
40	30.5052476893853\\
50	30.401384662685\\
60	30.1034975347506\\
};


\addplot [ forget plot, color=mygreen,line width=1.5 pt, mark size=3pt,   mark=triangle, mark options={line width=1.5pt, solid, rotate=270, mygreen}] table[row sep=crcr]{%
0	16.803027812658\\
10	31.8337315773867\\
20	31.9558298887739\\
30	31.9945698943112\\
40	31.9576537492116\\
50	31.8334115921837\\
60	31.6022854487592\\
};

\addplot [ forget plot, color=mycolor6,line width=1.5 pt, mark size=3.5pt, mark=10-pointed star, mark options={line width=1pt, solid, mycolor6}]  table[row sep=crcr]{%
0	18.8030184573641\\
10	33.8438594032587\\
20	33.9816784435654\\
30	34.0268273710397\\
40	33.982590709907\\
50	33.8436996475976\\
60	33.6024309289338\\
};


\addlegendimage{mark size=2.5pt, only marks, line width=1.8pt, mark=o, mark options={line width=1.5pt, solid, mycolor1}} 
\addlegendentry{D-FOC-U}


\addlegendimage{mark size=3pt, only marks, line width=1.8pt, mark=diamond, mark options={line width=1.5pt, solid, mycolor3}} 
\addlegendentry{D-MMSE-U}

\addlegendimage{mark size=2.5pt, only marks, line width=1.8pt, mark=triangle, mark options={line width=1.5pt, solid, mycolor4}} 
\addlegendentry{D-PEB-U}


\addlegendimage{mark size=3.5pt, only marks, line width=1.5pt, mark=10-pointed star, mark options={line width=1pt, solid, mycolor6}} 
\addlegendentry{ND-EIG-W}

\addlegendimage{mark size=2.5pt, only marks, line width=1.8pt, mark=triangle, mark options={line width=1.5pt, solid, rotate=270, mygreen}} 
\addlegendentry{ND-MMSE-U}

\end{axis}
\end{tikzpicture}%
}
            \caption*{(a)}
        \end{minipage}
        \hfill
        \begin{minipage}[t]{0.48\textwidth}
            \centering
            \hspace{-4mm}\resizebox{\linewidth}{!}{



\definecolor{mycolor1}{rgb}{0.00000,0.44700,0.74100}%
\definecolor{mycolor2}{rgb}{0.85000,0.32500,0.09800}%
\definecolor{mycolor3}{rgb}{0.92900,0.69400,0.12500}%
\definecolor{mycolor4}{rgb}{0.49400,0.18400,0.55600}%
\definecolor{mycolor5}{rgb}{0.46600,0.67400,0.18800}%
\definecolor{mycolor6}{rgb}{0.63500,0.07800,0.18400}%
\definecolor{mycolor7}{rgb}{0.30196,0.74510,0.93333}%
\definecolor{darkgray}{rgb}{0.4,0.4,0.4}
\definecolor{mygreen}{rgb}{0.0, 0.6, 0.3} 


\begin{tikzpicture}
\begin{axis}[%
    width=\columnwidth,
    width=0.9\columnwidth,
    scale only axis,
  xmin=-1,
    xmax=61,
    xtick={0,10, 20, 30, 40, 50, 60},
    ymin=1.5,
    ymax=35,
    ytick={5,10,15, 20, 25, 30, 35},
    xlabel style={font=\fontsize{12}{14}\selectfont, color=black},
    xlabel={$\Delta \phi$ [deg.]},
    ylabel style={font=\fontsize{12}{14}\selectfont, color=black},
    ylabel={$\Gamma$ [bps/Hz]},
      title={$d = 35 $ m},
    axis background/.style={fill=white},
  axis lines=box, 
    line width=0.6pt, 
xmajorgrids=false,
ymajorgrids=false,
    tick label style={font=\fontsize{12}{14}\selectfont},
    label style={font=\fontsize{12}{14}\selectfont},
    title style={font=\fontsize{12}{14}\selectfont},
    legend style={
    at={(1,0)},
    anchor=south east,
    legend cell align=left,
    align=left,
    draw=white!15!black,
    font=\fontsize{8}{10}\selectfont
},
axis lines=box,
xtick pos=bottom,
ytick pos=left,
xticklabel pos=lower,
yticklabel pos=left,
extra x ticks={},
extra y ticks={},
extra x tick style={draw=none},
extra y tick style={draw=none}
]

\addplot [ forget plot, color=mycolor1,line width=1.5pt, mark size=3pt, mark=o, mark options={line width=1.5pt, solid, mycolor1}]  table[row sep=crcr]{%
0	2.2636618181415\\
10	12.5862887635878\\
20	20.7513775694806\\
30	20.8202054839931\\
40	15.4408331612414\\
50	20.3243331624999\\
60	16.7613136287635\\
};

\addplot [ forget plot, color=mycolor1, dashed,line width=1.5pt, mark size=3pt, mark=o, mark options={line width=1.5pt, solid, mycolor1}]  table[row sep=crcr]{%
0	13.5157084825903\\
10	22.8403869850486\\
20	22.8637942664624\\
30	22.8948373810011\\
40	22.7055118274079\\
50	22.70729905162\\
60	21.9289738598957\\
};



\addplot [ forget plot, color=mycolor3,line width=1.5pt,mark size=3.5pt,   mark=diamond, mark options={line width=1.5pt, solid, mycolor3}]  table[row sep=crcr]{%
0	2.26366592899728\\
10	12.8783937518141\\
20	19.8740412409584\\
30	20.8885137717765\\
40	16.3101576876249\\
50	20.2433261046695\\
60	17.7314122088635\\
};

\addplot [ forget plot, color=mycolor3, dashed,line width=1.5pt,mark size=3pt,   mark=diamond, mark options={line width=1.5pt, solid, mycolor3}]  table[row sep=crcr]{%
0	13.5157075252633\\
10	22.8937080767669\\
20	22.9590007842534\\
30	22.9964569085417\\
40	22.8167267567317\\
50	22.7723351304406\\
60	21.9527643082231\\
};

\addplot [ forget plot, color=mycolor4,line width=1.5pt, mark size=3pt,   mark=triangle, mark options={line width=1.5pt, solid, mycolor4}]  table[row sep=crcr]{%
0	1.99916242998584\\
10	16.7931991027143\\
20	18.4416069839113\\
30	18.7318212898027\\
40	18.7927350195931\\
50	18.2240670666129\\
60	18.1701332681815\\
};

\addplot [ forget plot, color=mycolor4, dashed,line width=1.5pt, mark size=3pt,   mark=triangle, mark options={line width=1.5pt, solid, mycolor4}] table[row sep=crcr]{%
0	13.4629183338218\\
10	23.5703117929715\\
20	23.3830515201755\\
30	23.4239963784622\\
40	23.3706613602862\\
50	23.237222966145\\
60	22.9998572334872\\
};


\addplot [ forget plot, color=mygreen,line width=1.5pt, mark size=3pt,   mark=triangle, mark options={line width=1.5pt, solid, rotate=270, mygreen}] table[row sep=crcr]{%
0	13.1960481911452\\
10	24.6160353493635\\
20	24.7369821330996\\
30	24.7744202841605\\
40	24.7380716722152\\
50	24.6162833197643\\
60	24.3889183125207\\
};

\addplot [ forget plot, color=mycolor6,line width=1.5pt, mark size=3.5pt,   mark=10-pointed star, mark options={line width=1pt, solid,  mycolor6}]  table[row sep=crcr]{%
0	15.1959329580224\\
10	26.6256708966564\\
20	26.7614694005316\\
30	26.8054677532198\\
40	26.7620144785977\\
50	26.6257949433983\\
60	26.3886263143434\\
};


\addlegendimage{mark size=2.5pt, only marks, line width=1.8pt, mark=o, mark options={line width=1.5pt, solid, mycolor1}} 
\addlegendentry{D-FOC-U}


\addlegendimage{mark size=3pt, only marks, line width=1.8pt, mark=diamond, mark options={line width=1.5pt, solid, mycolor3}} 
\addlegendentry{D-MMSE-U}

\addlegendimage{mark size=2.5pt, only marks, line width=1.8pt, mark=triangle, mark options={line width=1.5pt, solid, mycolor4}} 
\addlegendentry{D-PEB-U}


\addlegendimage{mark size=3.5pt, only marks, line width=1.8pt, mark=10-pointed star, mark options={line width=1pt, solid, mycolor6}} 
\addlegendentry{ND-EIG-W}

\addlegendimage{mark size=2.5pt, only marks, line width=1.8pt, mark=triangle, mark options={line width=1.5pt, solid, rotate=270, mygreen}} 
\addlegendentry{ND-MMSE-U}

\end{axis}
\end{tikzpicture}%
}
            \caption*{(b)}
        \end{minipage}
    }
\caption{Sum-rate $\Gamma$ as a function of the angular separation $\Delta\phi$ between the two \acp{UE} for different \ac{AT-RIS} configuration strategies. Two propagation regimes are considered: (a) radiative near-field and (b) far-field. Solid lines represent the proposed configurations, while dashed lines denote their optimized counterparts, obtained by initializing the algorithm in~\cite{choi2024wmmse} with the corresponding configuration.}
      \label{fig:1A_sumrate}
\end{figure*}

Finally, the diagonal, focusing-based configurations, namely D-FOC-U and D-FOC-W, are also used as initializations for an optimization-based refinement according to the alternating optimization procedure presented in \cite{choi2024wmmse}. Specifically, the \ac{AMAF} beamforming matrix $\boldB$ is first optimized via the \ac{WMMSE} approach~\cite{christensen2008weighted}, followed by a gradient-based optimization of the \ac{T-RIS} configuration matrix $\boldsymbol{\Phi}$. This iterative optimization process gradually enhances the system sum-rate by alternately improving the active and passive beamforming components until convergence to a high-quality local optimum is achieved.

\subsection{Impact of Angular Separation Between UEs}\label{subsec:num_res_1}

To evaluate the \ac{AT-RIS} capability to spatially separate multiple \acp{UE} and its impact on system performance, we consider a two-\ac{UE} scenario ($K=2$) operating in the radiative near-field of the \ac{T-RIS}. The first \ac{UE} is positioned at a fixed angular location $\boldsymbol{\Theta}^{(1)} = \left[60^{\circ}, 0^{\circ}\right]$, at a distance of $d^{(1)} = 10\,\mathrm{m}$ from the \ac{T-RIS} reference point. The second \ac{UE} moves along a circular trajectory of radius $10\,\mathrm{m}$ within the $xy$-plane, such that the azimuthal separation $\Delta\phi$ between the two \acp{UE} is swept from $0^\circ$ to $60^\circ$. To assess performance across different propagation regimes, the same angular sweep is replicated in the far field by placing the circular trajectory at a distance of $35\,\mathrm{m}$ from the \ac{T-RIS}. This setup enables an accurate comparison between near-field and far-field propagation regimes, thus isolating the effect of angular separation on spatial multiplexing and system throughput.

Fig.~\ref{fig:1A_perUErate} illustrates the per-\ac{UE} rate $\gamma_k$ as a function of $\Delta\phi$ for three representative configurations in the near-field scenario at $d=10\,\mathrm{m}$, namely D-FOC-U, D-MMSE-U, and ND-EIG-W. Notably, diagonal \ac{T-RIS}-based configurations, such as D-FOC-U and D-MMSE-U, exhibit more oscillatory rate profiles compared to their non-diagonal counterpart. This behavior is primarily attributed to the limited spatial \ac{DoF} provided by diagonal \ac{T-RIS} phase profiles, which hinder the formation of directive beams with suppressed sidelobes. Nevertheless, both diagonal schemes demonstrate competitive performance, effectively suppressing inter-\ac{UE} interference even under limited angular separation. Particularly significant is the performance of D-FOC-U. Despite its simplicity, i.e., requiring only \ac{UE} location information for phase compensation and avoiding complex matrix operations, it achieves rates comparable to the more demanding D-MMSE-U scheme. The latter, in contrast, necessitates full complex-valued \ac{CSI} for all \acp{UE} and involves a computationally intensive matrix inversion, leading to significantly higher overhead. Furthermore, the plot reveals that even modest separations, e.g., $\Delta\phi = 10^\circ$, are sufficient for spatial discrimination, highlighting the practical relevance of D-FOC-U as a low-complexity yet effective solution.

This trend is further validated by the corresponding sum-rate curves reported in Fig.~\ref{fig:1A_sumrate}, which also extends the comparison to include the D-PEB-U and ND-MMSE-U strategies. Two representative propagation regimes are considered: (\textit{a}) a short-range configuration at $d=10\,\mathrm{m}$, representative of the radiative near-field, and (\textit{b}) a long-range setting at $d=35\,\mathrm{m}$, representative of the far-field. As expected, all configurations exhibit reduced performance in the far-field due to increased path loss. In the near-field case, D-FOC-U and D-MMSE-U achieve sum-rate trends that are qualitatively similar and quantitatively close to those of ND-EIG-W, confirming that properly configured diagonal \ac{T-RIS} architectures can deliver near-optimal performance. Additionally, the performance gap between optimized and non-optimized variants remains modest, especially in the far field. The D-PEB-U strategy, while achieving satisfactory performance at moderate angular separations, does not provide meaningful gains over simpler diagonal schemes and degrades at larger $\Delta\phi$ due to edge effects and suboptimal aperture utilization stemming from its partitioning approach.
Furthermore, the ND-MMSE-U configuration achieves sum-rate performance that closely approaches that of the ideal ND-EIG-W scheme, despite operating under non-cooperative assumptions. This suggests that well-designed non-diagonal \ac{T-RIS} transformations can approach theoretical limits without requiring receiver-side cooperation, offering a favorable balance between complexity and performance. Ultimately, in the far-field regime, all configurations converge to similar performance levels, indicating that the advantage of non-diagonal architectures diminishes as the channel becomes less spatially selective. In light of these results, the simple and scalable D-FOC-U scheme emerges as a compelling solution that ensures robust performance across both near- and far-field conditions.

\subsection{Effect of Power Allocation Strategy at the AMAF}\label{subsec:num_res_3}

\begin{figure}[t!]
    \centering
    \hspace{-7mm}\resizebox{\linewidth}{!}{\input{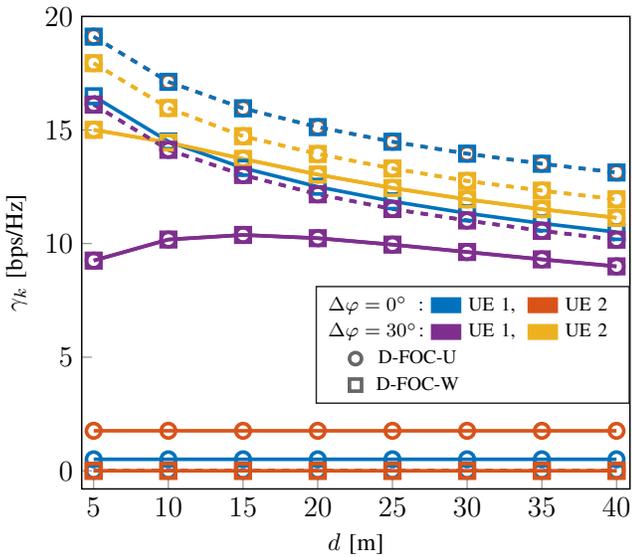}}  
\caption{Per-\ac{UE} achievable rate $\gamma_k$ for increasing link distances under D-FOC configuration, with either uniform or waterfilling power allocation and \acp{UE} spacing $\Delta\phi = \{0^\circ, 30^\circ\}$. Solid lines indicate the proposed scheme; dashed lines its optimized counterpart obtained via algorithm~\cite{choi2024wmmse}.}
    \label{fig:2A_waterfill}
\end{figure}

We now examine a fixed geometric setup in which the angular positions of the two \acp{UE} remain constant while the link distance is progressively increased from $5$ to $40$ meters, thereby transitioning from the near- to the far-field regime. Specifically, \ac{UE} 1 is placed at $\boldsymbol{\Theta}^{(1)} = \left[60^{\circ}, 0^{\circ}\right]$, while \ac{UE} 2 is either co-located with the first ($\Delta\phi = 0^\circ$) or placed at $\Delta\phi = 30^\circ$ in the \ac{AT-RIS} antenna boresight direction. 
 
Fig.~\ref{fig:2A_waterfill} shows the resulting per-\ac{UE} rate $\gamma_k$ achieved by the D-FOC configuration under two different power allocation strategies at the \ac{AMAF}, namely uniform (D-FOC-U) and waterfilling (D-FOC-W), both with and without the additional refinement provided by the algorithm in \cite{choi2024wmmse}. Several key insights emerge from the analysis. First, all configurations exhibit a gradual performance degradation as the distance increases, due to higher path loss and the reduced spatial resolution associated with far-field propagation. Second, the benefits of waterfilling are most evident in the very near-field, particularly when \acp{UE} are extremely close, i.e., where mutual interference among \acp{UE} severely constrains the system’s ability to maintain spatial separability. In this case, D-FOC-U fails to adequately resolve the inter-\ac{UE} interference, resulting in low rates for both \acp{UE}. Conversely, the D-FOC-W strategy significantly improves rate performance thanks to its adaptive power allocation, which enables a more efficient exploitation of the limited spatial \ac{DoF} available at short ranges. Conversely, when the angular spacing increases to $\Delta\phi = 30^\circ$, the performance gap between D-FOC-U and D-FOC-W narrows considerably, indicating that angular separation itself becomes the dominant factor in mitigating inter-\ac{UE} interference. This selective behavior confirms the usefulness of adaptive power allocation in interference-limited scenarios, especially when the angular spacing is below $10^\circ$. However, as soon as a modest angular separation is introduced, the benefits of waterfilling become marginal. In this case, the simple D-FOC-U approach, which does not require any form of \ac{CSI} beyond the \ac{UE}-to-\ac{T-RIS} phase compensation, performs comparably to its more complex waterfilling counterpart across all distances. Finally, in the far-field region, i.e., $d > 27\,\mathrm{m}$, all configurations converge toward similar rate levels, confirming that the effectiveness of adaptive power allocation vanishes as the channel becomes less sensitive to spatial shaping. These findings suggest that, for scalable and energy-efficient deployments, a uniform power allocation combined with a phase-compensating diagonal \ac{T-RIS} offers an attractive balance between complexity and performance, particularly when a minimum angular spacing among \acp{UE} is ensured.

\subsection{Effect of Link Distance and Propagation Regime}\label{subsec:num_res_2}

We now analyze how the link distance and the associated propagation regime (near-field vs. far-field) influence system performance. As previously noted, Fig.~\ref{fig:1A_sumrate} reveals a consistent trend: all considered strategies experience a decline in sum-rate as the average distance of the \acp{UE} increases, primarily due to path loss. Among them, non-diagonal architectures such as ND-EIG-W and ND-MMSE-U achieve the highest sum-rates in the very near-field but also suffer the steepest decline, with losses approaching $10\,\mathrm{bps/Hz}$ as link distance increases. In contrast, diagonal \ac{T-RIS} configurations exhibit a more gradual degradation, highlighting their greater robustness against propagation-induced impairments. Fig.~\ref{fig:1A_sumrate} also underscores the relevance of azimuthal \ac{UE} separation. Even modest angular spacing, i.e., approximately $10^\circ$ for non-diagonal schemes and $20^\circ$ for diagonal ones, substantially enhances spatial separability and leads to more stable performance. This emphasizes the critical role of angular diversity in mitigating inter-\ac{UE} interference, particularly in dense near-field deployments.

\begin{figure}[t!]
    \centering
   \hspace{-5mm} \resizebox{\linewidth}{!}{\input{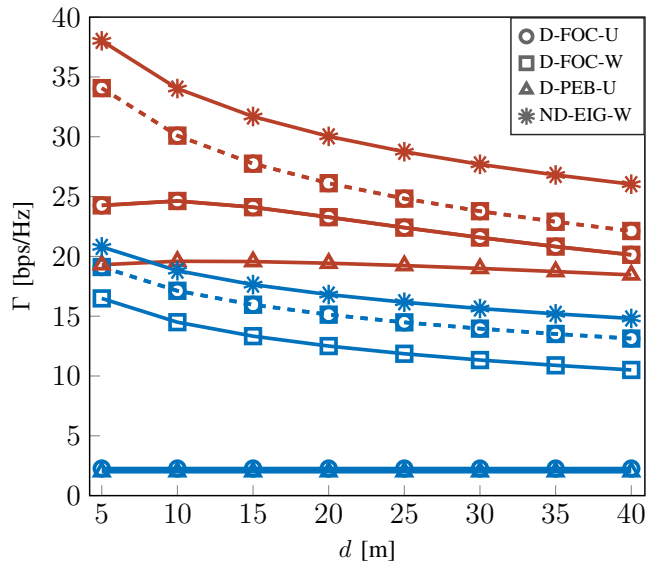}}  
\caption{Sum rate $\Gamma$ as a function of the link distance for different \ac{AT-RIS} configuration strategies. Blue curves correspond to a \ac{UE} spacing of $\Delta\phi = 0^\circ$, while dark red curves correspond to the $\Delta\phi = 30^\circ$ case. Solid lines indicate the proposed scheme, while dashed lines indicate its optimized counterpart via algorithm~\cite{choi2024wmmse}.}
\label{fig:2A_sumrate}
\end{figure}

To further examine this behavior, we refer to the same setup as in Fig.~\ref{fig:2A_waterfill}, where both \acp{UE} progressively increase their distance from the \ac{T-RIS} while maintaining a fixed angular position for \ac{UE} 1. Fig.~\ref{fig:2A_sumrate} shows the corresponding sum-rate achieved by four representative configurations, i.e., D-FOC-U, D-FOC-W, D-PEB-U, and ND-EIG-W, under two angular separation scenarios: $\Delta\phi = 0^\circ$ (blue curves) and $\Delta\phi = 30^\circ$ (dark red curves). Solid lines represent the baseline version of each scheme, while dashed lines indicate their refined counterparts through algorithm \cite{choi2024wmmse}.
In the $\Delta\phi = 0^\circ$ setting, diagonal schemes such as D-FOC-U and D-PEB-U yield near-zero sum-rates unless supported by additional optimization or adaptive power allocation at the feeder side. This behavior stems from the absence of spatial separability, which hampers interference suppression and power focusing. In contrast, D-FOC-W maintains non-trivial performance even in such a limiting case. When further optimized, it achieves sum-rates that closely approach those of the more complex ND-EIG-W strategy, while requiring substantially lower computational complexity and system overhead. As the angular separation increases to $\Delta\phi = 30^\circ$, the impact of the propagation regime becomes more evident. Schemes that perform well in the near-field suffer marked performance degradation as the distance grows and the system's operation transitions into the far-field regime. This effect is especially pronounced for ND-EIG-W, whose performance converges toward that of simpler diagonal approaches. These findings confirm that diagonal architectures, when paired with proper feeder-side power allocation and spatial focusing mechanisms, can achieve an effective trade-off between spectral efficiency, computational simplicity, and robustness across a broad range of deployment conditions. Such versatility makes them well-suited for practical implementations, especially in scenarios with dynamic \ac{UE} distributions and partial or outdated channel knowledge.

\begin{figure*}[t!]
    \centering
    \begin{minipage}[t]{0.48\textwidth}
        \centering
        \hspace{-3mm}\resizebox{\linewidth}{!}{


\definecolor{mycolor1}{rgb}{0.00000,0.44700,0.74100}%
\definecolor{mycolor2}{rgb}{0.85000,0.32500,0.09800}%
\definecolor{mycolor3}{rgb}{0.92900,0.69400,0.12500}%
\definecolor{mycolor4}{rgb}{0.49400,0.18400,0.55600}%
\definecolor{mycolor5}{rgb}{0.46600,0.67400,0.18800}%
\definecolor{mycolor6}{rgb}{0.30100,0.74500,0.93300}%
\definecolor{mybrick}{rgb}{0.722,0.251,0.157} 
\definecolor{mygreen}{rgb}{0.0, 0.6, 0.3} 
\definecolor{darkgray}{rgb}{0.4,0.4,0.4}            

\begin{tikzpicture}
\begin{axis}[%
    width=\columnwidth,
    width=0.9\columnwidth,
    scale only axis,
xmin=1.6,
xmax=16.3,
    xtick={2, 4, 6, 8, 10, 12, 14, 16},
ymin=-0.85,
ymax=17,
    ytick={0,2,4,6,8,10,12,14,16},
    xlabel style={font=\fontsize{12}{14}\selectfont, color=black},
    xlabel={$K$},
    ylabel style={font=\fontsize{12}{14}\selectfont, color=black},
ylabel={$\mathrm{E}[\gamma_k]\ \mathrm{[bps/Hz]}$},
    title={},
    axis background/.style={fill=white},
  axis lines=box, 
    line width=0.6pt, 
xmajorgrids=false,
ymajorgrids=false,
    tick label style={font=\fontsize{12}{14}\selectfont},
    label style={font=\fontsize{12}{14}\selectfont},
    title style={font=\fontsize{12}{14}\selectfont},
    legend style={
    at={(1,1)},
    anchor=north east,
    legend cell align=left,
    align=left,
    draw=white!15!black,
    fill=white,
    font=\fontsize{8}{10}\selectfont
},
axis lines=box,
xtick pos=bottom,
ytick pos=left,
xticklabel pos=lower,
yticklabel pos=left,
extra x ticks={},
extra y ticks={},
extra x tick style={draw=none},
extra y tick style={draw=none}
]

\addplot [color=mycolor1, line width=1.5pt, mark size=3pt, mark=o, mark options={line width=1.5pt,solid, mycolor1}]
  table[row sep=crcr]{%
2	8.97987088647012\\
3	6.25467361362404\\
4	4.81777872767427\\
5	3.99579391804541\\
6	3.38452265141114\\
7	2.87657063865993\\
8	2.45287074036203\\
9	2.13441023550645\\
10	1.9394388195153\\
11	1.6992125922425\\
12	1.54086739089871\\
13	1.35107229892219\\
14	1.21906913671748\\
15	1.09345782445593\\
16	0.994335462168853\\
};
\addlegendentry{D-FOC-U}


\addplot [color=mycolor3, line width=1.5pt, mark size=3.5pt, mark=diamond, mark options={line width=1.5pt,solid, mycolor3}]
  table[row sep=crcr]{%
2	8.75913215300901\\
3	6.16249714936711\\
4	4.46933248292467\\
5	3.57880589307544\\
6	2.84443708895022\\
7	2.3840038388607\\
8	2.02001086855479\\
9	1.63458797937369\\
10	1.38854976372379\\
11	1.11606109596644\\
12	0.967045560270185\\
13	0.789711639012835\\
14	0.662913258763162\\
15	0.539476454725947\\
16	0.456282424820128\\
};
\addlegendentry{D-MMSE-U}

\addplot [color=mycolor4, line width=1.5pt, mark size=3pt, mark=triangle, mark options={line width=1.5pt,solid, mycolor4}]
  table[row sep=crcr]{%
2	8.49337139299365\\
3	4.26724656621482\\
4	5.80990846910362\\
5	1.3168732476934\\
6	2.91427880228941\\
7	0.645225597454788\\
8	1.46176363907757\\
9	1.96522534846876\\
10	0.81762680831744\\
11	0.250110804528926\\
12	0.995582252154779\\
13	0.190594426979367\\
14	0.363662670313129\\
15	0.555629358175966\\
16	0.630378316922629\\
};
\addlegendentry{D-PEB-U}

\addplot [color=mybrick, line width=1.5pt, mark size=4pt, mark=10-pointed star, mark options={line width=1pt,solid, mybrick}]
  table[row sep=crcr]{%
2	15.194177359766\\
3	14.5794996590203\\
4	14.1512216953325\\
5	13.6980514334081\\
6	13.374024364498\\
7	13.0518626543884\\
8	12.7225362364905\\
9	12.4505688806878\\
10	12.1141661541161\\
11	11.8379790523514\\
12	11.5610600071709\\
13	11.3295044361316\\
14	11.0030910819349\\
15	10.6790368739685\\
16	10.3945068974907\\
};
\addlegendentry{ND-EIG-W}

\addplot [color=mygreen, line width=1.5pt, mark size=3pt, mark=triangle, mark options={line width=1.5pt,solid, rotate=270, mygreen}]
  table[row sep=crcr]{%
2	13.7767437994628\\
3	12.3676860419751\\
4	11.0916942460361\\
5	10.2512242340484\\
6	9.18480179085555\\
7	8.51651768254417\\
8	7.71939216303506\\
9	7.02122561159471\\
10	6.00896648036555\\
11	5.23640889327172\\
12	4.5881118877995\\
13	3.980233757469\\
14	3.33628195281592\\
15	2.89227057919183\\
16	2.37321215802943\\
};
\addlegendentry{ND-MMSE-U}

\end{axis}

\end{tikzpicture}%
}
        \caption*{\hspace{10mm}(a)}
    \end{minipage}%
    \hfill
    \begin{minipage}[t]{0.48\textwidth}
        \centering
        \hspace{-3mm}\resizebox{\linewidth}{!}{


\definecolor{mycolor1}{rgb}{0.00000,0.44700,0.74100}%
\definecolor{mycolor2}{rgb}{0.85000,0.32500,0.09800}%
\definecolor{mycolor3}{rgb}{0.92900,0.69400,0.12500}%
\definecolor{mycolor4}{rgb}{0.49400,0.18400,0.55600}%
\definecolor{mycolor5}{rgb}{0.46600,0.67400,0.18800}%
\definecolor{mycolor6}{rgb}{0.30100,0.74500,0.93300}%
\definecolor{mybrick}{rgb}{0.722,0.251,0.157} 
\definecolor{mygreen}{rgb}{0.0, 0.6, 0.3} 
\definecolor{darkgray}{rgb}{0.4,0.4,0.4}            

\begin{tikzpicture}
\begin{axis}[%
    width=\columnwidth,
    width=0.9\columnwidth,
    scale only axis,
xmin=1.6,
xmax=16.3,
    xtick={2, 4, 6, 8, 10, 12, 14, 16},
ymin=-0.85,
ymax=17,
    ytick={0,2,4,6,8,10,12,14,16, 18},
    xlabel style={font=\fontsize{12}{14}\selectfont, color=black},
    xlabel={$K$},
    ylabel style={font=\fontsize{12}{14}\selectfont, color=black},
ylabel={$\mathrm{Var}[\gamma_k]\ \mathrm{[bps^2/Hz^2]}$},
    title={},
    axis background/.style={fill=white},
  axis lines=box, 
    line width=0.6pt, 
xmajorgrids=false,
ymajorgrids=false,
    tick label style={font=\fontsize{12}{14}\selectfont},
    label style={font=\fontsize{12}{14}\selectfont},
    title style={font=\fontsize{12}{14}\selectfont},
    legend style={
    at={(0,1)},
    anchor=north west,
    legend cell align=left,
    align=left,
    draw=white!15!black,
    fill=white,
    font=\fontsize{8}{10}\selectfont
},
axis lines=box,
xtick pos=bottom,
ytick pos=left,
xticklabel pos=lower,
yticklabel pos=left,
extra x ticks={},
extra y ticks={},
extra x tick style={draw=none},
extra y tick style={draw=none}
]
\addplot [color=mycolor1, line width=1.5pt, mark size=3pt, mark=o, mark options={solid, mycolor1}]
  table[row sep=crcr]{%
2	9.43928876703061\\
3	6.5006769211596\\
4	5.20794788703989\\
5	3.73836751899293\\
6	3.20377104863818\\
7	2.47332309357697\\
8	1.99564701990694\\
9	1.58480849552769\\
10	1.29870204412434\\
11	1.05784948562115\\
12	0.901171155108652\\
13	0.717352404121192\\
14	0.629331374191953\\
15	0.544830810948479\\
16	0.518778696195391\\
};
\addlegendentry{D-FOC-U}


\addplot [color=mycolor3, line width=1.5pt, mark size=3.5pt, mark=diamond, mark options={solid, mycolor3}]
  table[row sep=crcr]{%
2	9.17230515660494\\
3	6.74883710859324\\
4	5.83518638411987\\
5	4.50014807990416\\
6	3.75957942761606\\
7	2.97345675010593\\
8	2.45410649705166\\
9	1.8931095728198\\
10	1.52412780287038\\
11	1.19265494135987\\
12	0.982059326907994\\
13	0.736891590820694\\
14	0.587348201618856\\
15	0.44190311495588\\
16	0.382501759748818\\
};
\addlegendentry{D-MMSE-U}

\addplot [color=mycolor4, line width=1.5pt, mark size=3pt, mark=triangle, mark options={solid, mycolor4}]
  table[row sep=crcr]{%
2	5.99325355876205\\
3	2.53208537954304\\
4	7.03094209413187\\
5	0.420087763304158\\
6	2.31988120740399\\
7	0.103950839962508\\
8	0.588678720874687\\
9	1.275647409555\\
10	0.222362086530523\\
11	0.0230660170257218\\
12	0.34774147280666\\
13	0.0266459119803455\\
14	0.0459077995661351\\
15	0.121758104823522\\
16	0.144426636185107\\
};
\addlegendentry{D-PEB-U}

\addplot [color=mybrick, line width=1.5pt, mark size=4pt, mark=10-pointed star, mark options={line width=1pt, solid,mybrick}]
  table[row sep=crcr]{%
2	3.23329586504237\\
3	3.96853244816887\\
4	3.81797980534281\\
5	5.05427517527319\\
6	5.71066098289466\\
7	6.04076300292202\\
8	7.26941547045208\\
9	8.18891042691822\\
10	9.17181810203682\\
11	10.0717048213958\\
12	11.1969418960474\\
13	11.6050220416648\\
14	13.2999283747093\\
15	14.8005220640621\\
16	16.4337562825066\\
};
\addlegendentry{ND-EIG-W}

\addplot [color=mygreen, line width=1.5pt, mark size=3pt, mark=triangle, mark options={solid, rotate=270, mygreen}]
  table[row sep=crcr]{%
2	2.44520193934599\\
3	2.68642604798256\\
4	4.72552560534783\\
5	5.39675485941891\\
6	7.61494381357898\\
7	7.20362563999959\\
8	8.6729086323711\\
9	9.09590606065507\\
10	9.32494134742421\\
11	9.57831467138548\\
12	9.09446729044541\\
13	8.14577557511168\\
14	7.44822401368119\\
15	6.57974379691319\\
16	5.73317499620516\\
};
\addlegendentry{ND-MMSE-U}

\end{axis}

\end{tikzpicture}%
}
        \caption*{\hspace{10mm}(b)}
    \end{minipage}

    \vspace{3mm}

    \begin{minipage}[t]{0.48\textwidth}
        \centering
        \hspace{-5mm}\resizebox{\linewidth}{!}{


\definecolor{mycolor1}{rgb}{0.00000,0.44700,0.74100}%
\definecolor{mycolor2}{rgb}{0.85000,0.32500,0.09800}%
\definecolor{mycolor3}{rgb}{0.92900,0.69400,0.12500}%
\definecolor{mycolor4}{rgb}{0.49400,0.18400,0.55600}%
\definecolor{mycolor5}{rgb}{0.46600,0.67400,0.18800}%
\definecolor{mycolor6}{rgb}{0.30100,0.74500,0.93300}%
\definecolor{mycolor2}{rgb}{0.722,0.251,0.157} 

\definecolor{mycolor7}{rgb}{0.0, 0.6, 0.3} 
\definecolor{darkgray}{rgb}{0.4,0.4,0.4}            

\begin{tikzpicture}
\begin{axis}[%
    width=\columnwidth,
    width=0.9\columnwidth,
    scale only axis,
xmin=1.6,
xmax=16.3,
    xtick={2, 4, 6, 8, 10, 12, 14, 16},
ymin=-1.5,
ymax=170,
    ytick={0,20, 40, 60, 80, 100, 120, 140 ,160, 180},
    xlabel style={font=\fontsize{12}{14}\selectfont, color=black},
    xlabel={$K$},
    ylabel style={font=\fontsize{12}{14}\selectfont, color=black},
    ylabel={$\Gamma$ [bps/Hz]},
    title={},
    axis background/.style={fill=white},
  axis lines=box, 
    line width=0.6pt, 
 xmajorgrids=false,
ymajorgrids=false,
    tick label style={font=\fontsize{12}{14}\selectfont},
    label style={font=\fontsize{12}{14}\selectfont},
    title style={font=\fontsize{12}{14}\selectfont},
    legend style={
    at={(0,1)},
    anchor=north west,
    legend cell align=left,
    align=left,
    draw=white!15!black,
    fill=white,
    font=\fontsize{8}{10}\selectfont
},
axis lines=box,
xtick pos=bottom,
ytick pos=left,
xticklabel pos=lower,
yticklabel pos=left,
extra x ticks={},
extra y ticks={},
extra x tick style={draw=none},
extra y tick style={draw=none}
]

\addplot [color=mycolor1,  line width=1.5pt, mark size=3pt, mark=o, mark options={line width=1.5pt,solid, mycolor1}]
  table[row sep=crcr]{%
2	17.9597417729402\\
3	18.7640208408721\\
4	19.2711149106971\\
5	19.9789695902271\\
6	20.3071359084668\\
7	20.1359944706195\\
8	19.6229659228962\\
9	19.209692119558\\
10	19.394388195153\\
11	18.6913385146675\\
12	18.4904086907845\\
13	17.5639398859885\\
14	17.0669679140447\\
15	16.4018673668389\\
16	15.9093673947016\\
};
\addlegendentry{D-FOC-U}


\addplot [color=mycolor3,  line width=1.5pt, mark size=4pt, mark=diamond, mark options={line width=1.5pt,solid, mycolor3}]
  table[row sep=crcr]{%
2	17.518264306018\\
3	18.4874914481013\\
4	17.8773299316987\\
5	17.8940294653772\\
6	17.0666225337013\\
7	16.6880268720249\\
8	16.1600869484383\\
9	14.7112918143632\\
10	13.8854976372379\\
11	12.2766720556308\\
12	11.6045467232422\\
13	10.2662513071668\\
14	9.28078562268427\\
15	8.09214682088921\\
16	7.30051879712205\\
};
\addlegendentry{D-MMSE-U}

\addplot [color=mycolor4,  line width=1.5pt, mark size=3pt, mark=triangle, mark options={line width=1.5pt,solid, mycolor4}]
  table[row sep=crcr]{%
2	16.9867427859873\\
3	12.8017396986445\\
4	23.2396338764145\\
5	6.58436623846701\\
6	17.4856728137365\\
7	4.51657918218352\\
8	11.6941091126206\\
9	17.6870281362188\\
10	8.1762680831744\\
11	2.75121884981819\\
12	11.9469870258574\\
13	2.47772755073177\\
14	5.0912773843838\\
15	8.33444037263949\\
16	10.0860530707621\\
};
\addlegendentry{D-PEB-U}

\addplot [color=mycolor2,  line width=1.5pt, mark size=4pt, mark=10-pointed star, mark options={line width=1pt,solid, mycolor2}]
  table[row sep=crcr]{%
2	30.388354719532\\
3	43.738498977061\\
4	56.6048867813302\\
5	68.4902571670405\\
6	80.2441461869883\\
7	91.3630385807185\\
8	101.780289891924\\
9	112.05511992619\\
10	121.141661541161\\
11	130.217769575866\\
12	138.73272008605\\
13	147.283557669711\\
14	154.043275147089\\
15	160.185553109527\\
16	166.312110359852\\
};
\addlegendentry{ND-EIG-W}

\addplot [color=mycolor7,  line width=1.5pt, mark size=3pt, mark=triangle, mark options={line width=1.5pt,solid, rotate=270, mycolor7}]
  table[row sep=crcr]{%
2	27.5534875989256\\
3	37.1030581259255\\
4	44.3667769841443\\
5	51.2561211702422\\
6	55.1088107451333\\
7	59.6156237778092\\
8	61.7551373042804\\
9	63.1910305043523\\
10	60.0896648036555\\
11	57.6004978259889\\
12	55.057342653594\\
13	51.743038847097\\
14	46.707947339423\\
15	43.3840586878775\\
16	37.9713945284709\\
};
\addlegendentry{ND-MMSE-U}

\end{axis}
\end{tikzpicture}%
}
        \caption*{\hspace{10mm}(c)}
    \end{minipage}%
    \hfill
    \begin{minipage}[t]{0.48\textwidth}
        \centering
        \hspace{-5mm}\resizebox{\linewidth}{!}{


\definecolor{mycolor1}{rgb}{0.00000,0.44700,0.74100}%
\definecolor{mycolor2}{rgb}{0.85000,0.32500,0.09800}%
\definecolor{mycolor3}{rgb}{0.92900,0.69400,0.12500}%
\definecolor{mycolor4}{rgb}{0.49400,0.18400,0.55600}%
\definecolor{mycolor5}{rgb}{0.46600,0.67400,0.18800}%
\definecolor{mycolor6}{rgb}{0.30100,0.74500,0.93300}%
\definecolor{mybrick}{rgb}{0.722,0.251,0.157} 
\definecolor{mygreen}{rgb}{0.0, 0.6, 0.3} 
\definecolor{darkgray}{rgb}{0.4,0.4,0.4}            

\begin{tikzpicture}
\begin{axis}[%
    width=\columnwidth,
    width=0.9\columnwidth,
    scale only axis,
xmin=1.6,
xmax=16.3,
    xtick={2, 4, 6, 8, 10, 12, 14, 16},
ymin=0.42,
ymax=1.03,
ytick={0.4, 0.5, 0.6, 0.7, 0.8, 0.9, 1.0},
    xlabel style={font=\fontsize{12}{14}\selectfont, color=black},
    xlabel={$K$},
    ylabel style={font=\fontsize{12}{14}\selectfont, color=black},
    ylabel={$\mathcal{J}(\{\gamma_k\})$},
    title={},
    axis background/.style={fill=white},
  axis lines=box, 
    line width=0.6pt, 
xmajorgrids=false,
ymajorgrids=false,
    tick label style={font=\fontsize{12}{14}\selectfont},
    label style={font=\fontsize{12}{14}\selectfont},
    title style={font=\fontsize{12}{14}\selectfont},
    legend style={
    at={(0,0)},
    anchor=south west,
    legend cell align=left,
    align=left,
    draw=white!15!black,
    fill=white,
    font=\fontsize{8}{10}\selectfont
},
axis lines=box,
xtick pos=bottom,
ytick pos=left,
xticklabel pos=lower,
yticklabel pos=left,
extra x ticks={},
extra y ticks={},
extra x tick style={draw=none},
extra y tick style={draw=none}
]

\addplot [color=mycolor1, line width=1.5pt, mark size=3pt, mark=o, mark options={line width=1.5pt,solid, mycolor1}]
  table[row sep=crcr]{%
2	0.975703504520817\\
3	0.902455075024482\\
4	0.852966530513846\\
5	0.837423128743388\\
6	0.807970023575452\\
7	0.792289452583841\\
8	0.772659364235639\\
9	0.763831177753624\\
10	0.763135887881137\\
11	0.753547744143951\\
12	0.745584143280866\\
13	0.738266891484834\\
14	0.723141998285326\\
15	0.707751039298574\\
16	0.675414904767261\\
};
\addlegendentry{D-FOC-U}


\addplot [color=mycolor3, line width=1.5pt, mark size=3.5pt, mark=diamond, mark options={line width=1.5pt,solid, mycolor3}]
  table[row sep=crcr]{%
2	0.967288458566203\\
3	0.894357838519267\\
4	0.824217683101231\\
5	0.789645328488363\\
6	0.745824186274711\\
7	0.715163289098872\\
8	0.689577624949256\\
9	0.655321896575544\\
10	0.637771202443907\\
11	0.602875843444172\\
12	0.575517835086287\\
13	0.552895207573976\\
14	0.524370375381315\\
15	0.488878108987438\\
16	0.447039100425116\\
};
\addlegendentry{D-MMSE-U}

\addplot [color=mycolor4, line width=1.5pt, mark size=3pt, mark=triangle, mark options={line width=1.5pt,solid, mycolor4}]
  table[row sep=crcr]{%
2	0.992676052458254\\
3	0.911932014959224\\
4	0.860692849766228\\
5	0.842815345671986\\
6	0.807493999346899\\
7	0.826291161187048\\
8	0.804265291631804\\
9	0.770030264002038\\
10	0.778048038913587\\
11	0.751443656705567\\
12	0.75750763924327\\
13	0.599599905080481\\
14	0.761142017482924\\
15	0.737842258694541\\
16	0.747201946379476\\
};
\addlegendentry{D-PEB-U}

\addplot [color=mybrick, line width=1.5pt, mark size=4pt, mark=10-pointed star, mark options={line width=1pt,solid, mybrick}]
  table[row sep=crcr]{%
2	0.991241569547528\\
3	0.985317148885337\\
4	0.984184925611981\\
5	0.9761670147588\\
6	0.970998928142075\\
7	0.967383487758424\\
8	0.958508935381808\\
9	0.951209928641995\\
10	0.942424994311758\\
11	0.934185791900838\\
12	0.923779402269573\\
13	0.918118374336463\\
14	0.901984618328533\\
15	0.886101517557056\\
16	0.868924965478108\\
};
\addlegendentry{ND-EIG-W}

\addplot [color=mygreen, line width=1.5pt, mark size=3pt, mark=triangle, mark options={line width=1.5pt,solid, rotate=270, mygreen}]
  table[row sep=crcr]{%
2	0.999999976381299\\
3	0.999999939395978\\
4	0.999993058506692\\
5	0.99900510122326\\
6	0.998722452389945\\
7	0.994174236104648\\
8	0.987678719200654\\
9	0.976022643628014\\
10	0.953613484844808\\
11	0.921068583553723\\
12	0.886835566134814\\
13	0.854520184430731\\
14	0.793964832853235\\
15	0.744399369122744\\
16	0.676965676495731\\
};
\addlegendentry{ND-MMSE-U}

\end{axis}
\end{tikzpicture}%
}
        \caption*{\hspace{10mm}(d)}
    \end{minipage}

    \caption{System performance in terms of (a) mean per-UE rate $\mathrm{E}[\gamma_k]$, (b) standard deviation of per-UE rates $\mathrm{Var}[\gamma_k]$, (c) total sum-rate $\Gamma$, and (d) Jain fairness index $\mathcal{J}(\{\gamma_k\})$ as a function of the number of \acp{UE} for different \ac{AT-RIS} configuration strategies.}
    \label{fig:3_scalability}
\end{figure*}

\subsection{System Scalability: Impact of UEs Number}\label{subsec:num_res_4}

To evaluate the scalability of the proposed near-field \ac{MU} system, we consider a scenario in which the number of single-antenna \acp{UE} increases from $K = 2$ to $K = 16$. For each $K$, the \acp{UE} are randomly and independently positioned within a circular sector oriented towards the \ac{T-RIS}. The link distances and angular positions for each \ac{UE} $k \in \mathcal{K}$ are drawn from the ranges $d^{(k)} \in [5, 30]\,\mathrm{m}$ and $\phi^{(k)} \in [30^\circ, 150^\circ]$, respectively. This random deployment is repeated over $N_{\mathrm{MC}} = 1000$ independent Monte Carlo trials, and performance metrics are averaged to ensure statistical reliability.
In this regard, Fig.~\ref{fig:3_scalability} displays (a) the average per-\ac{UE} rate $\mathrm{E}[\gamma_k]$, (b) the corresponding rate variance $\mathrm{Var}[\gamma_k]$, (c) the total sum-rate $\Gamma$, and (d) the Jain fairness index $\mathcal{J}(\{\gamma_k\})$ as a function of the number of served \acp{UE}, for several \ac{AT-RIS} configuration strategies. This joint analysis reveals key trade-offs between spectral efficiency, fairness, and robustness under increasing \ac{UE} density.

As illustrated in Fig.~\ref{fig:3_scalability}(c), the ND-EIG-W benchmark consistently provides the highest sum-rate, exhibiting a quasi-linear increase in $\Gamma$ up to $K=16$. However, this gain comes at the cost of growing rate disparity, as reflected in the increasing variance in Fig.~\ref{fig:3_scalability}(b). Still, the Jain index reported in Fig.~\ref{fig:3_scalability}(d) remains relatively high, indicating acceptable fairness even in denser regimes with large $K$. This reflects the typical behavior of aggressive spatial multiplexing strategies that exploit all available channel \ac{DoF}. While effective in \ac{SU}-\ac{MIMO} settings, where energy is focused along the dominant eigenmode, this approach yields different results in \ac{MU}-\ac{MISO} systems without any explicit \ac{UE} fairness constraints. Specifically, the ND-EIG-W scheme greedily allocates resources to spatial directions with stronger channel energy, favoring well-aligned \acp{UE} and penalizing those with weaker coupling or reduced spatial orthogonality. As a result, spectral efficiency is maximized at the expense of fairness, which limits the scheme’s suitability in practical \ac{MU} deployments lacking \ac{UE}-aware optimization. 
A similar trend is observed for ND-MMSE-U, which achieves good performance for $K \leq 10$, but suffers sharp fairness degradation as $K$ increases. As shown in Fig.~\ref{fig:3_scalability}(d), ND-MMSE-U yields the highest fairness for $K \leq 8$, but rapidly deteriorates beyond this point, highlighting limited scalability in high \ac{UE} density regimes. Conversely, ND-EIG-W maintains higher fairness for $K > 10$, confirming its robustness despite increased rate variance.
Interestingly, D-FOC-U and ND-MMSE-U exhibit nearly identical fairness at $K = 16$, demonstrating the ability of focusing via a diagonal \ac{T-RIS} to sustain high fairness even under heavy \ac{UE} load. This highlights the strength of D-FOC strategies in maintaining scalable and stable performance when paired with suitable feeder-side beamforming and power control.
Among all configurations, D-FOC-U emerges as the most balanced solution, sustaining an average per-\ac{UE} rate above $1\,\mathrm{bps/Hz}$ with minimal rate dispersion, even at $K = 16$. This favorable behavior stems from two interrelated factors. First, the symmetry of the \ac{T-RIS} beamforming structure ensures that each \ac{UE} is served with equal-gain focalized beams, unlike ND-EIG-W, which instead biases transmission toward dominant directions. Second, the eigenvalue decay in the \ac{AMAF} near-field channel matrix becomes more pronounced with increasing $K$, revealing saturation of the effective spatial \ac{DoF}. This leads to non-uniform power allocation among the feeder-side beamformers and limits the ability to maintain equal-quality links. Such limitations are intrinsic to near-field propagation and can only be partially mitigated by enlarging the apertures at the \ac{T-RIS} and the \ac{AMAF}. Consequently, rate degradation occurs homogeneously across \acp{UE}, counterintuitively promoting fairness and stability.
The D-MMSE-U scheme exhibits a similar trend across all metrics, mainly due to its inability to capture inter-\ac{UE} correlation under diagonal \ac{T-RIS} constraints, leading to inefficient use of spatial resources. Finally, D-PEB-U shows the most irregular and non-monotonic behavior. While it occasionally achieves favorable metrics due to specific geometrical arrangements of the \acp{UE}, its performance is generally unreliable and highly sensitive to their spatial distribution, leading to significant variability across deployment instances and a lack of robustness in dense settings. As $K$ increases, the aperture allocated to each \ac{UE} shrinks due to fixed-sector partitioning, resulting in severe resolution loss. This confirms that PEB-based schemes are unsuitable for dense \ac{MU} settings unless each \ac{UE} is served by an independent \ac{T-RIS} segment, an impractical solution for large $K$.

In summary, while non-diagonal \ac{T-RIS} schemes such as ND-EIG-W and ND-MMSE-U provide excellent throughput in moderate-load regimes, their scalability is limited. Diagonal strategies like D-FOC-U offer a more robust and balanced trade-off between efficiency, \acp{UE} fairness, and complexity, making them suitable for realistic \ac{MU} scenarios with large \ac{UE} populations.

\section{Conclusions}\label{sec:Conclusions}

This work investigates a modular beamforming framework for \ac{AT-RIS}-enabled near-field \ac{MU}-\ac{MISO} systems, in which beamforming tasks are split between power allocation at the \ac{AMAF} and \ac{UE}-specific transmissive precoding at the \ac{T-RIS}. This separation allows the two stages to be independently configured and jointly optimized, enabling multiple architectures with distinct complexity-performance trade-offs. Numerical results reveal that non-diagonal strategies such as ND-EIG-W yield high sum-rate in near-field regimes, but exhibit fairness limitations and degraded scalability under increased \ac{UE} density or far-field conditions. In contrast, diagonal \ac{T-RIS} schemes, as the proposed D-FOC-U, achieve robust and scalable performance with low complexity, leveraging only phase compensation and requiring minimal \ac{CSI}. The analysis further emphasizes the impact of angular separation in mitigating inter-\ac{UE} interference, and shows that adaptive power allocation strategies (e.g., waterfilling) are effective primarily when \acp{UE} are closely spaced. Moreover, it is confirmed that eigenmode-based transmission, although optimal in \ac{SU}-\ac{MIMO} settings, becomes unbalanced in \ac{MU}-\ac{MISO} scenarios due to its inherently greedy nature. This behavior stems from the fact that eigenmode-based transmission combined with waterfilling power allocation aims to maximize the overall sum-rate under the implicit assumption of \ac{UE} cooperation, thus often favoring strong channels while penalizing weaker \ac{UE} channels in non-cooperative \ac{MU} contexts.
Overall, the proposed diagonal \ac{AT-RIS} architectures with focusing-based precoding and uniform power allocation emerge as a practical and scalable solution for near-field \ac{MU}-\ac{MISO}, offering a favorable trade-off between complexity, fairness, and performance. Future research could explore the integration of \ac{UE}-specific quality-of-service constraints and the application of learning-based methods to further improve adaptability and efficiency under dynamic network conditions.

\section*{Acknowledgments}

G. Torcolacci and D. Dardari are with the Department of Electrical, Electronic, and Information Engineering “Guglielmo Marconi”, University of Bologna, Italy, and CNIT-WiLab, Bologna, Italy (e-mail: \{g.torcolacci, davide.dardari\}@unibo.it). \\
M. Schellmann is with Huawei Technologies German Research Center, Munich, Germany (e-mail: malte.schellmann@huawei.com).\\
This work was partially supported by the European Union under the Italian National Recovery and Resilience Plan (NRRP) of NextGenerationEU, partnership on “Telecommunications of the Future” (PE00000001 - program “RESTART”) and by the EU Horizon project TIMES (Grant no. 101096307). G. Torcolacci was funded by an NRRP Ph.D. grant.\\
\textit{This work has been submitted to IEEE for possible publication. Copyright may be transferred without notice.}

\bibliographystyle{IEEEtran}
\bibliography{references}

\end{document}